\documentclass[a4paper]{article}
\usepackage{fullpage}
\usepackage{amsmath,amsthm,amssymb,amsfonts,mathrsfs,eucal,amsthm}
\usepackage{graphicx,hyperref}
\usepackage{dsfont}
\usepackage[stable]{footmisc}

\def\states{\mathfrak{S}}

\def\vphi{{\varphi}}
\def\lle{\scriptscriptstyle{\leq}}
\def\={\scriptscriptstyle{=}}

\def\Tr{\operatorname{Tr}}

\def\>{\rangle}
\def\<{\langle}
\def\sH{\mathscr{H}}
\def\M{\mathscr{M}}

\def\cS{\mathcal{S}}
\def\sA{\mathscr{A}}

\def\leq{\leqslant}
\def\be{\begin{equation}}
\def\ee{\end{equation}}
\def\bea{\begin{eqnarray}}
\def\eea{\end{eqnarray}}
\def\ben{\begin{eqnarray*}}
\def\een{\end{eqnarray*}}

\def\N#1{\left|\!\left|{#1}\right|\!\right|}

\def\openone{\mathds{1}}

\def\tr{\operatorname{Tr}} 

\def\eps{\varepsilon}

\def\I{\widetilde{I}}

\def\tQ{\widetilde{Q}}

\def\P{\mathfrak{p}}

\def\reff#1{(\ref{#1})}

\def\B{\mathfrak{b}}

\def\eE{\mathfrak{E}}
\def\mM{\mathcal{M}}

\renewcommand{\ge}{\geqslant}
\renewcommand{\le}{\leqslant}

\renewcommand{\qedsymbol}{\nobreak \ifvmode \relax \else
      \ifdim\lastskip<1.5em \hskip-\lastskip
      \hskip1.5em plus0em minus0.5em \fi \nobreak
      \vrule height0.75em width0.5em depth0.25em\fi}

\newtheorem{theorem}{Theorem}

\newtheorem{lemma}{Lemma}

\theoremstyle{remark}
\newtheorem{remark}{Remark}

\theoremstyle{definition}
\newtheorem{definition}{Definition}

\begin{document}

\title{General theory of environment-assisted entanglement distillation}

\author{Francesco Buscemi\footnote{Institute for Advanced
    Research, University of Nagoya, Chikusa-ku, Nagoya 464-8601, Japan
    (e-mail:buscemi@iar.nagoya-u.ac.jp)} \and Nilanjana
    Datta\footnote{Statistical Laboratory, University of Cambridge,
    Wilberforce Road, Cambridge CB3 0WB, UK
    (e-mail:n.datta@statslab.cam.ac.uk)}}

\date{\today}

\maketitle
\begin{abstract}
  We evaluate the one-shot entanglement of assistance for an arbitrary
  bipartite state. This yields another interesting result, namely
  a characterization of the one-shot distillable entanglement of a 
  bipartite pure state.   This result is shown to be stronger than 
  that obtained by specializing the one-shot hashing bound to pure states. 
  Finally, we show how the one-shot result yields the operational 
  interpretation of the asymptotic entanglement of assistance proved 
  in [Smolin et  al. Phys. Rev. A {\bf 72}, 052317 (2005)]. 
\end{abstract}

\section{Introduction}

One of the most basic and widely studied entanglement measures for
bipartite quantum states is the \emph{entanglement of formation}
(EoF)~\cite{eof}, a quantity so named because it was intended to
quantify the resources needed to create (or form) a given bipartite
entangled state. The EoF of any bipartite pure state is quantified by
the entropy of entanglement, which is equal to the von Neumann entropy
of the reduced state of a subsystem.  The EoF of a bipartite mixed
state $\rho_{AB}$, is then defined via the convex roof extension, that
is, as the minimum average entanglement of an ensemble of pure states
that represents $\rho_{AB}$:
\begin{equation}
  \label{eq:10}
  E_F(\rho_{AB}):=\min_{\eE}\sum_ip_iS(\rho^i_A),
\end{equation}
where $\eE=\{p_i,|\psi^i_{AB}\>\}$ is an ensemble of pure biparite
states such that $\sum_ip_i|\psi^i\>\<\psi^i|=\rho_{AB}$, and
$S(\rho^i_A)$ is the von Neumann entropy of the reduced state
$\rho^i_A=\Tr_B|\psi^i\>\<\psi^i|_{AB}$.  The popularity of the EoF is
partly due to its formal elegance and the many nice properties it
enjoys \cite{haya,matthias}, and perhaps also due to its connections
with the additivity problem in quantum information theory
\cite{eof-add,hastings}.

From the operational point of view, the
EoF is associated with the entanglement
manipulation protocol by which two
distant parties, say Alice and Bob, prepare a given bipartite 
quantum state, starting from an initial entangled state which they 
share, by using only local operations and classical 
communication (LOCC). It turns out that the optimal
(i.e., minimum) rate, at which entanglement has to be consumed in order
for Alice and Bob to create multiple copies of the state with asymptotically
vanishing error, is given by the regularized EoF of the state~\cite{cost}.

Soon after the introduction of the EoF, another quantity, namely
the \emph{entanglement of assistance} (EoA)~\cite{eoa}, was introduced 
as its ``dual''. It is defined analogously to EoF but with the 
minimisation over ensembles replaced by a maximisation, i.e., 
\begin{equation}
  \label{eq:19}
  E_A(\rho_{AB}):=\max_\eE \sum_ip_iS(\rho^i_A).
\end{equation}

Unlike the EoF, the EoA is not an entanglement monotone and hence it
can in general increase under local operations and classical
communication~\cite{gilad-spekkens}. However, like the EoF, the EoA
too can be associated with an entanglement manipulation protocol,
namely the one by which Alice and Bob distill entanglement from an
initial mixed bipartite state which they share, when a third party
(say Charlie), who holds the purification of the state, assists them.
Charlie is allowed to do local operations on his share of the
tripartite pure state, and his assistance is in the form of one-way
classical communication to Alice and Bob.  This is the sort of
scenario which occurs, for example, in the case of
\emph{environment-assisted quantum error
  correction}~\cite{greg,hay-king,assistance,only-andreas,pavia,fb},
in which errors, incurred from sending quantum information through a
noisy environment, are corrected by using classical information
obtained from a measurement on the environment. In this case the
tripartite structure Alice-Bob-Charlie is mirrored by the structure
sender-receiver-environment, and the assistance from Charlie is
replaced by the ability to perform measurements on the environment and
to exploit the resulting information for error correction.

Another area in which the EoA arises, is in the study of
\emph{localizable entanglement} in spin
systems~\cite{local0,local1,local2,local3}. The scenario here is as
follows: a pure state of a system of $n\gg 1$ interacting spins is
given, and the goal is to localize (or ``focus'') as much entanglement
as possible between two arbitrarily chosen spins, by performing a
suitable measurement on the remaining $n-2$ spins. In this case, the
assisting party is actually divided into many subsystems (which are
the $n-2$ spins) and so it is natural to ask what happens when the
assisting measurements are restricted to be \emph{local} in each
subsystem. The amount of entanglement that can be focussed in this
case is referred to as the localizable entanglement, and it is always
at most as much as the EoA. In fact, in the case in which the
assisting parties are allowed to perform \emph{global} measurements on
all their subsystems at once, the localizable entanglement obviously
equals the EoA.

In the literature, one encounters cases in which the EoA is used to
characterize operational tasks of assisted distillation studied in the
generic scenario, where no assumptions are made on the state to be
distilled.  This is often referred to as the ``one-shot'' scenario.
However, the definition of the EoA given in eq.~(\ref{eq:19}) has been
shown to have an operational relevance only in the asymptotic regime,
i.e., when asymptotically many copies of the same state are available
for assisted distillation~\cite{assistance}. This points to an
apparent mismatch between the operational task and the quantity used
to characterize it. In order to remedy this problem, one should start
from the operational task itself, and from it, \emph{evaluate} an
expression quantifying the amount of entanglement that can be
distilled under assistance from a single sample of an arbitrary
bipartite state. This leads to a \emph{one-shot} EoA, which, by its
very construction, has a direct operational interpretation.

In this paper, we obtain bounds on the one-shot EoA in the scenario
mentioned above. As an intermediate step, we obtain a complete
characterization of the one-shot distillable entanglement of an
arbitrary bipartite pure state. This result improves on previous known
bounds, derived from the one-shot hashing
bound~\cite{distil}. Finally, we apply our results to get an
alternative proof of the fact~\cite{assistance} that the regularized
EoA is the optimal rate of environment-assisted entanglement
distillation in the asymptotic scenario.

The paper is organized as follows. In Section~2 we introduce the
necessary notation and definitions. In Section~3 we evaluate the
one-shot distillable entanglement of a pure bipartite state. The
one-shot entanglement of assistance is introduced in Section~4 and
evaluated in Section~5. Section~6 deals with the asymptotic scenario,
where some previous results are recovered. Finally, Section~7
concludes the paper with a summary and an open question.

\section{Notation and definitions}
\label{prelim}
\subsection{Mathematical preliminaries}
Let ${\cal B}(\sH)$ denote the algebra of linear operators acting on a
finite--dimensional Hilbert space $\sH$ and let $\states(\sH) \subset
{\cal B}(\sH)$ denote the subset of positive operators of unit trace
(states). Further, let $\openone\in {\cal B}(\sH)$ denote the
identity operator.  Throughout this paper we restrict our
considerations to finite-dimensional Hilbert spaces, and we take the
logarithm to base $2$. For any given pure state $|\phi\>$, we denote
the projector $|\phi\>\<\phi|$ simply as $\phi$. Moreover, for any
state $\rho$, we define $\Pi_\rho$ to be the projector onto the
support of $\rho$.

For a state $\rho\in\states(\sH)$, the von Neumann entropy is defined
as $S(\rho):=-\tr\rho\log\rho$. Further, for a state $\rho$ and a
positive operator $\sigma$ such that ${\rm{supp }} \rho \subseteq
{\rm{supp }}\sigma$, the quantum relative entropy is defined as
$S(\rho||\sigma) = \tr \rho \log \rho - \rho \log \sigma,$ whereas the
relative R\'enyi entropy of order $\alpha \in (0,1)$ is defined as
\be\label{relren} S_\alpha (\rho || \sigma) := \frac{1}{\alpha - 1}
\log\bigl[ \tr(\rho^\alpha \sigma^{1-\alpha})\bigr].  \ee
For given orthonormal bases $\{|i_A\rangle\}_{i=1}^d$ and
$\{|i_B\rangle\}_{i=1}^d$ in isomorphic Hilbert spaces
$\sH_A\simeq\sH_B$ of dimension $d$, we define the standard maximally
entangled state (MES) of rank $M \le d$ to be
\begin{equation}\label{MES-M}
|\Psi^M_{AB}\>= \frac{1}{\sqrt{M}} \sum_{i=1}^M |i_A\rangle\otimes |i_B\rangle.
\end{equation}

In order to measure how close two states are, we will use the
fidelity, defined as
\begin{equation}\label{fidelity-aaa}
F(\rho, \sigma):= \tr \sqrt{\sqrt{\rho} \sigma \sqrt{\rho}}
=\N{\sqrt{\rho}\sqrt{\sigma}}_1,
\end{equation}
and the trace distance
\begin{equation}\label{trace-aaa}
 \N{\rho-\sigma}_1:=\Tr|\rho-\sigma|.
\end{equation}
In what follows, equations~(\ref{fidelity-aaa}) and~(\ref{trace-aaa})
will sometimes be directly extended to operators other than normalized
states, if required.

The trace distance between two states $\rho$ and $\sigma$ is related
to the fidelity $F(\rho, \sigma)$ as follows (see e.~g.~\cite{nielsen}):
\begin{equation}
  1-F(\rho,\sigma) \leq \frac{1}{2} \N{\rho -
    \sigma}_1 \leq \sqrt{1-F^2(\rho, \sigma)},
\label{fidelity}
\end{equation}
where we use the notation $F^2(\rho, \sigma) = \bigl(F(\rho,\sigma)
\bigr)^2$

The following lemmas will prove useful.

\begin{lemma}[\cite{bowen-datta}]\label{bowen}
  For any self-adjoint operators $A$ and $B$, and any positive
  operator $0\le P\le\openone$,
\begin{equation}\nonumber
\Tr[P(A-B)]\le\Tr(A-B)_+\le \N{A-B}_1,
\end{equation}
where $(A-B)_+$ denotes the positive part of the difference operator
$A-B$.
\end{lemma}

\begin{lemma}[Gentle measurement lemma~\cite{winter99,ogawanagaoka02}]
  \label{gmlemma}
  For a state $\rho\in\states(\sH)$ and an operator $0\le
  \Lambda\le\openone$, if $\Tr(\rho\ \Lambda) \ge 1 - \delta$, then
$$\N{\rho -   {\sqrt{\Lambda}}\rho{\sqrt{\Lambda}}}_1 \le {2\sqrt{\delta}}.$$
The same holds if $\rho$ is a subnormalized density operator.
\end{lemma}

\begin{lemma}\label{fid3}
For any pure state $|\phi\rangle$ and any given $\eps\ge 0$,
if $0\le P\le \openone$ is an operator such that $\tr (P\phi) \ge 1-\eps$,
then 
\begin{equation}
F( \sqrt{P}|\phi\rangle, |\phi\rangle) \ge 1 - \sqrt{\eps}.
\end{equation}
\end{lemma}
\begin{proof}
Since, $\tr (P\phi) \ge 1-\eps$, by Lemma \ref{gmlemma} we have that 
$$\|\sqrt{P} \phi \sqrt{P} - \phi \|_1 \le 2 \sqrt{\eps}.$$
The lower bound on the trace distance in \reff{fidelity} then yields 
\begin{equation}
  F( \sqrt{P}|\phi\rangle, |\phi\rangle) \equiv F(\sqrt{P} \phi \sqrt{P},\phi)\ge 1 - \sqrt{\eps}.
\end{equation}
\end{proof}

\begin{lemma}\label{lemma:accessory}
  For any normalized state $\rho$ and any $0\le P\le\openone$, if
  $\Tr[P\rho]\ge 1-\eps$, then
\begin{equation}
  F(\omega,\rho)\ge 1-2\sqrt{\eps},
\end{equation}
where $\omega:=\frac{\sqrt{P}\rho\sqrt{P}}{\Tr[P\rho]}$.
\end{lemma}

\begin{proof}
  By Lemma~\ref{gmlemma}, the condition $\Tr[P\rho]\ge 1-\eps$ implies
  that $\N{\sqrt{P}\rho\sqrt{P}-\rho}_1\le 2\sqrt{\eps}$. Let us
  define $\tilde\omega:=\sqrt{P}\rho\sqrt{P}$. Due to Lemma~11
  in~\cite{distil}, we have that
  \begin{equation}
\begin{split}
  F(\tilde\omega,\rho):&=\N{\sqrt{\tilde\omega}\sqrt{\rho}}_1\\
  &\ge\frac{\Tr[P\rho]+1}2 - \frac 12\N{\tilde\omega-\rho}_1\\
  & \ge 1-\frac{\eps}2-\sqrt{\eps}\\
  &\ge 1-2\sqrt{\eps}.
\end{split}
\end{equation}
Let $\omega$ be the normalized state defined as
$\omega:=\frac{\tilde\omega}{\Tr(\tilde\omega)}$. Since
$F(\omega,\rho)\ge F(\tilde\omega,\rho)$, we obtain the statement of
the lemma.
\end{proof}

In this paper we consider entanglement distillation under LOCC
transformations.  In this context, a result by Lo and Popescu
\cite{lopopescu} on entanglement manipulation of bipartite {\em{pure
    states}} plays a crucial role.  They proved that any LOCC
transformation ($AB \mapsto A'B'$) on a bipartite pure state
$|\phi_{AB}\rangle$, shared between two distant parties Alice and Bob,
is equivalent to a LOCC transformation with only one-way classical
communication, which can be represented as follows: \be\label{form}
\Lambda(\phi_{AB}) = \sum_k (U_k \otimes E_k)\phi_{AB} (U_k \otimes
E_k)^\dagger,\ee where the operators $U_k$ are unitary and the
operators $E_k$ satisfy the relation $\sum_k E_k^\dagger E_k =
\openone_{B}$. Henceforth, we say that an LOCC transformation is of
the {\em{Lo-Popescu form}} if it can be expressed as in
(\ref{form}). Consequently, for a map $\Lambda$ of the Lo-Popescu
form, we have \bea
\Lambda (\openone_A \otimes \sigma_B) &=& \sum_k U_kU_k^\dagger \otimes E_k\sigma_B E_k^\dagger,\nonumber\\
&=& \openone_{A'} \otimes \tau_{B'},
\label{lopop}
\eea
where $\tau_{B'}:= \sum_k  E_k\sigma_B E_k^\dagger$.
. 
%
\subsection{Entropies and coherent
  information}\label{entropies}

Optimal rates of the entanglement distillation protocols considered 
in this paper are expressible in terms
of the following entropic quantities:
\smallskip

For any $\rho,\sigma\ge 0$,
any $0\le P\le\openone$, and any $\alpha\in(0,\infty)\backslash\{1\}$,
we define the following entropic function (introduced in \cite{qcap})
\begin{equation}\label{quasi-ent}
S_\alpha^P(\rho\|\sigma):=\frac{1}{\alpha-1}\log\Tr[\sqrt{P}\rho^\alpha\sqrt{P}\sigma^{1-\alpha}].
\end{equation}
Notice that, for $P=\openone$, the function defined above reduces to
the relative R\'enyi entropy of order $\alpha$ given by \reff{relren}.

In this paper, we are in particular interested in the quantity,
\begin{equation}\label{eq:asda}
S_0^P(\rho\|\sigma):=\lim_{\alpha\searrow 0}S_\alpha^P(\rho\|\sigma)= 
-\log\Tr[\sqrt{P}\Pi_\rho\sqrt{P}\ \sigma],
\end{equation}
where $\Pi_\rho$ denotes the projector onto the support of $\rho$. 

Note that
\be
S_0^{\openone}(\rho\|\sigma) = S_0(\rho\|\sigma) := - \log (\tr \Pi_\rho \sigma),
\ee
which is the relative R\'enyi entropy of order zero. This quantity acts
as a parent quantity for the {\em{zero-coherent information}}, defined as
follows:
\begin{equation}\label{zero-coh}
  I^{A\to B}_{0}(\rho_{AB}):=\min_{\sigma_B\in\states(\sH_B)}S_0(\rho_{AB}\|\openone_A\otimes\sigma_B),
\end{equation}
the nomenclature arising from its analogy with the ordinary coherent
information $I^{A\to B}(\rho_{AB})$, which is expressible in a similar
manner, when the zero-relative R\'enyi entropy is replaced by the
ordinary relative entropy: \bea
\label{coh}
I^{A\to
  B}(\rho_{AB})&:=& S(\rho_B)-S(\rho_{AB})\label{coh11}\\
&\equiv &
\min_{\sigma_B\in\states(\sH_B)}S(\rho_{AB}\|\openone_A\otimes\sigma_B).
\eea The above equality follows easily by expanding the last term
according to the definition of the quantum relative entropy and by
noticing that the minimum is achieved when $\sigma_B=\rho_B$, since
$\log(\openone_A\otimes\sigma_B)=\openone_A\otimes\log\sigma_B$ and
$S(\rho_B\|\sigma_B)\ge 0$. (For the complete derivation see, for
example, Lemma~6 in Ref.~\cite{qcap}.)

If $\Psi^\rho_{ABE}$ is a purification of the state $\rho_{AB}$,
then \be\label{neg} I^{A\to B}(\rho_{AB}) = - I^{A\to E}(\rho_{AE}),
\ee where $\rho_{AE}= \tr_B \Psi^\rho_{ABE}$.

Note in particular that for a MES of rank $M$, as defined by
\reff{MES-M}, 
\begin{equation}
I^{A\to B}_{0}(\Psi^M_{AB}) = I^{A\to B}(\Psi^M_{AB})= \log M.
\end{equation}

Another entropic quantity of relevance in this paper is the
{\em{min-entropy of a state}}, which is defined for any state $\rho$
as follows~\cite{renner-phd}:
\be S_{\min}(\rho) = - \log
\bigl[\lambda_{\max}(\rho)\bigr],
\ee
where $\lambda_{\max}(\rho)$ denotes the maximum eigenvalue of the
state $\rho$.

For one-shot entanglement distillation protocols it is natural to
allow for a finite accuracy, i.e., a non-zero error (say $\eps \ge
0$), in the extraction of singlets from a given state. In this case
the optimal rates of the protocols are given by ``smoothed versions''
of the entropic quantities introduced above. In order to define them
we consider the following sets of positive operators for any
normalized state $\rho$, and any $\eps >0$:
\begin{equation}
  \B(\rho;\eps):=\left\{\sigma:\sigma\ge 0,\ \Tr[\sigma]=1,\ F^2(\rho,\sigma)\ge1-\eps^2\right\},\label{ball}
\end{equation}
\begin{equation}
  \P(\rho;\eps):=\left\{P:0\le P\le\openone,\ \Tr[P\rho]\ge1-\eps\right\}.\label{P-ball}
\end{equation}

Further, by restricting the states $\sigma$ in \reff{ball} to be pure
states, we obtain the subset
\begin{equation}\label{b*}
  \B_*(\rho;\eps):=\left\{|\vphi\rangle:\,\vphi\in\B(\rho;\eps)\right\}.
\end{equation}
It was proved in \cite{marco} that for a bipartite pure state 
$|\phi_{AB}\rangle$, for any $\eps \ge 0$,
\begin{equation}\label{equi}
  \left\{\Tr_A[\vphi_{AB}]:\vphi_{AB}\in \B_*(\phi_{AB};\eps)\right\}= \B(\rho^\phi_B; \eps),
\end{equation}
where $\rho^{B}_\phi:= \Tr_A \phi_{AB}$.

The relevant smoothed entropic quantities are then defined as follows:
\begin{definition}
For any given $\eps \ge 0$ the {\em{smoothed}} min-entropy of a state $\rho$
is defined as
\be
S_{\min}^\eps(\rho) := \max_{\bar{\rho} \in \B(\rho;\eps)} S_{\min} ({\bar{\rho}}).
\label{smooth_min}
\ee
\end{definition}

We consider two different smoothed versions of the zero-coherent information, defined as
follows: 
\begin{definition} The state-smoothed zero-coherent information is given by
\begin{equation}\label{eq:i}
  I^{A\to B}_{0,\eps}(\rho_{AB}):=\max_{\bar\rho_{AB}\in \B(\rho_{AB};\eps)}\min_{\sigma_B\in\states(\sH_B)}S_0(\bar\rho_{AB}\|\openone_A\otimes\sigma_B),
\end{equation}
and the operator-smoothed zero-coherent information is given by
\begin{equation}\label{eq:itilda}
  \I_{0,\eps}^{A\to B}(\rho_{AB}):=\max_{P\in \P(\rho_{AB};\eps)}\min_{\sigma_B\in\states(\sH_B)}S_0^P(\rho_{AB}\|\openone_A\otimes\sigma_B).
\end{equation}
\end{definition}

\begin{remark}
  A variant of the operator-smoothing introduced above has been used
  in~\cite{WR,datta-brandao,datta-hsieh}. Note, however, that in this
  paper we only use the operator-smoothed zero-coherent information as
  an intermediate quantity: the main results are given entirely in
  terms of the more familiar state-smoothed quantities.
\end{remark}

The following technical lemmas involving the operator-smoothed coherent 
information are used in proving some of our main results.
\begin{lemma}\label{lemma1}
  If for a bipartite state $\rho_{AB}$ and a pure state
  $|\psi_{AB}\rangle$, for any given $\eps \ge 0$,
  \begin{equation}
    F^2(\rho_{AB},
    \psi_{AB})\ge 1-\eps,
\label{fid2}
\end{equation}
then
\be
 \I_{0,\eps}^{A\to B}(\rho_{AB}) \ge  I_{0}^{A\to B}(\psi_{AB}).
\ee
\end{lemma}
\begin{proof}
  Since the state $\psi_{AB}$ is pure, $F^2(\rho_{AB},
  \psi_{AB})=\tr[\rho_{AB} \psi_{AB}]\ge 1-\eps$. It follows that
  $\psi_{AB} \in \P(\rho_{AB};\eps)$. Using this fact,
  \reff{eq:itilda} and \reff{quasi-ent}, we obtain \bea
  \I_{0,\eps}^{A\to B}(\rho_{AB}) &\ge &
  \min_{\sigma_B\in\states(\sH_B)}
  \Bigl[- \log \tr\bigl({\sqrt{\psi_{AB}}}\Pi_{\rho_{AB}}{\sqrt{\psi_{AB}}}(\openone_A\otimes \sigma_B) \bigr)\Bigr]\nonumber\\
  &\ge & \min_{\sigma_B\in\states(\sH_B)}\Bigl[- \log \tr\bigl({\psi_{AB}}(\openone_A\otimes \sigma_B) \bigr)\Bigr]\nonumber\\
  &=& I_{0}^{A\to B}(\psi_{AB}).  \eea where the second inequality
  follows from the fact that
  ${\sqrt{\psi_{AB}}}\Pi_{\rho_{AB}}{\sqrt{\psi_{AB}}}\le{\psi_{AB}}$,
  since $\Pi_{\rho_{AB}}\le\openone_{AB}$.
\end{proof}
\begin{lemma}\label{lemma2}
  For any bipartite pure state $|\phi_{AB}\>$, any LOCC map
  $\Lambda:AB\mapsto A'B'$, and any $\eps\ge0$,
\begin{equation}\label{eq:lemma2-stat1}
  \I_{0,2\sqrt{\eps}}^{A\to B}(\phi_{AB}) \ge \I_{0,\eps}^{A'\to
    B'}(\Lambda(\phi_{AB})).
\end{equation}
\end{lemma}
\begin{proof}
Since the LOCC map $\Lambda$ acts on a pure state, without loss of generality 
we can assume it to be of the Lo-Popescu form (\ref{form}).
Defining $\omega_{A'B'} := \Lambda(\phi_{AB})$, we have, starting from \reff{eq:itilda},
\begin{equation}
\begin{split}
  \label{eq1.1} 
  \I_{0,\eps}^{A'\to B'}(\Lambda(\phi_{AB})) &= \max_{P\in
    \P(\omega_{A'B'};\eps)}\min_{\sigma_{B'}\in\states(\sH_{B'})}
  \left\{- \log \Tr \left[ \sqrt{P} \Pi_{\omega_{A'B'}} \sqrt{P}\ (\openone_{A'}\otimes\sigma_{B'})\right]\right\}\\
  &= \min_{\sigma_{B'}\in\states(\sH_{B'})} \left\{- \log \Tr \left[
      \sqrt{P_0} \Pi_{\omega_{A'B'}} \sqrt{P_0}\
      (\openone_{A'}\otimes\sigma_{B'})\right]\right\}\\
  &\le - \log \Tr \left[ \sqrt{P_0} \Pi_{\omega_{A'B'}} \sqrt{P_0}\ (\openone_{A'}\otimes{\tilde{\sigma}}_{B'})\right]\\
  &=  - \log \Tr \left[ \sqrt{P_0} \Pi_{\omega_{A'B'}} \sqrt{P_0}\ \Lambda(\openone_A\otimes\sigma_{B})\right]\\
  &= - \log \Tr \left[ \Lambda^*\left(\sqrt{P_0} \Pi_{\omega_{A'B'}}
      \sqrt{P_0}\right)\ (\openone_A\otimes\sigma_{B})\right],
  \end{split}
\end{equation}
for any state $\sigma_B \in \states(\sH_B)$. In the above, $P_0$ is
the operator in $\P(\omega_{A'B'};\eps)$ for which the maximum in the
first line is achieved; ${\tilde{\sigma}}_{B'}$ is a state in
$\states(\sH_{B'})$ such that
$\openone_{A'}\otimes{\tilde{\sigma}}_{B'}=\Lambda(\openone_A\otimes\sigma_{B})$,
and $\Lambda^*:A'B'\mapsto AB$ denotes the dual map of $\Lambda$,
defined, for any operator $X$ and state $\rho$, as $ \tr[X
\Lambda(\rho)] = \tr[\Lambda^* (X) \rho]$.

Let us now define $\tQ_{AB} := \Lambda^*(\sqrt{P_0}
\Pi_{\omega_{A'B'}} \sqrt{P_0})$. Then, continuing from 
equation~\reff{eq1.1}, we obtain 
\begin{equation}\label{eq:asdfgh}
\begin{split}
  \I_{0,\eps}^{A'\to B'}(\Lambda(\phi_{AB})) &\le - \log \Tr \left[ \tQ_{AB}\ (\openone_A\otimes\sigma_{B})\right]\\
&\le - \log \tr  \left[ \sqrt{\tQ_{AB}}\ \phi_{AB}\ \sqrt{\tQ_{AB}}\ (\openone_A \otimes \sigma_B)\right],
\end{split} 
\end{equation}
for any state $\sigma_{B}$, since
$\tQ_{AB}\ge\sqrt{\tQ_{AB}}\phi_{AB}\sqrt{\tQ_{AB}}$. 
Since the above inequality holds for any state $\sigma_B$, we have in
particular that
\begin{equation}\label{31}
   \I_{0,\eps}^{A'\to B'}(\Lambda(\phi_{AB}))\le \min_{\sigma_B} \left\{- \log \tr  \left[
    \sqrt{\tQ_{AB}}\ \phi_{AB}\ \sqrt{\tQ_{AB}}\ (\openone_A \otimes
    \sigma_B)\right]\right\}
\end{equation}

We next prove that
$\tQ_{AB}\in\P(\phi_{AB};2\sqrt{\eps})$. In fact, since $P_0 \in
\P(\omega_{A'B'};\eps)$, by the Gentle Measurement Lemma, 
\begin{equation}
  \N{\Lambda(\phi_{AB}) - \sqrt{P_0} \Lambda(\phi_{AB})\sqrt{P_0}}_1
  \le 2 \sqrt{\eps}.
\label{gm1}
\end{equation}
We therefore have, by definition of $\tQ_{AB}$, \bea\label{33} \tr
\left[ \tQ_{AB} \phi_{AB}\right] &=&
\tr \left[ \sqrt{P_0} \Pi_{\Lambda(\phi_{AB})} \sqrt{P_0}\ \Lambda(\phi_{AB})\right]\nonumber\\
&=& \tr \left[ \Pi_{\Lambda(\phi_{AB})} \sqrt{P_0}
  \Lambda(\phi_{AB}) \sqrt{P_0}\right]\nonumber\\
&=&\tr \left[ \Pi_{\Lambda(\phi_{AB})}\Lambda(\phi_{AB})\right] \nonumber\\
& & \quad + \tr \left[ \Pi_{\Lambda(\phi_{AB})}\bigl(\sqrt{P_0}\
  \Lambda(\phi_{AB}) \sqrt{P_0} - \Lambda(\phi_{AB})\bigr)\right]\nonumber\\
& \ge & 1 - \|\sqrt{P_0}\Lambda(\phi_{AB}) \sqrt{P_0} - \Lambda(\phi_{AB})\|_1 \nonumber\\
&\ge & 1- 2\sqrt{\eps}, \eea where the second line follows from the
cyclicity of the trace, the first inequality follows from
Lemma~\ref{bowen}, and the last inequality follows from (\ref{gm1}).
This implies that $\tQ_{AB} \in \P(\phi_{AB};2\sqrt{\eps})$. Hence, we
have from \reff{31}
\begin{equation}
\begin{split}
  \I_{0,\eps}^{A'\to B'}(\Lambda(\phi_{AB}))&\le \min_{\sigma_B}
  \left\{- \log \tr \left[ \sqrt{\tQ_{AB}}\ \phi_{AB}\
      \sqrt{\tQ_{AB}}\ (\openone_A \otimes
      \sigma_B)\right]\right\} \\
  & \le \max_{P \in \P(\phi_{AB};2\sqrt{\eps})}\min_{\sigma_B}\left\{ - \log
  \tr \left[ \sqrt{P}\ \phi_{AB}\ \sqrt{P}\ (\openone_A \otimes \sigma_B) \right]\right\}\\
  &\equiv\I_{0,2\sqrt{\eps}}^{A\to B}(\phi_{AB}),
\end{split}
\end{equation}
which completes the proof.
\end{proof}

\begin{lemma}
\label{lemma:opsm-stsm}
For any bipartite pure state $|\phi_{AB}\>$ and any $\eps\ge0$,
\begin{equation}\label{eq:lemma2-stat11}
 I_{0,\eps}^{A\to B}(\phi_{AB})\ge S_{\min}^{\eps}(\rho^\phi_A),
\end{equation}
where $\rho^\phi_A:=\Tr_B\phi_{AB}$. Further,
\begin{equation}\label{eq:lemma2-stat}
  \I_{0,\eps}^{A\to B}(\phi_{AB})\le S_{\min}^{{\eps'}}(\rho^\phi_A) -\log(1-\eps),
\end{equation}
where $\eps'=2\eps^{\frac 14}$.
\end{lemma}

\begin{proof}
  We first prove~(\ref{eq:lemma2-stat11}). Starting from~\reff{eq:i}
  we have:
  \begin{equation}
\begin{split}
  I_{0,\eps}^{A\to B}(\phi_{AB}) :&=\max_{\bar\rho_{AB}\in
    \B(\phi_{AB};\eps)}\min_{\sigma_B\in\states(\sH_B)}S_0(\bar\rho_{AB}\|\openone_A\otimes\sigma_B)\\
  &\ge \max_{\bar\vphi_{AB}\in
    \B_*(\phi_{AB};\eps)}\min_{\sigma_B\in\states(\sH_B)}S_0(\bar\vphi_{AB}\|\openone_A\otimes\sigma_B)\\
  &=\max_{\bar\vphi_{AB}\in
    \B_*(\phi_{AB};\eps)}\min_{\sigma_B\in\states(\sH_B)}\left\{-\log\Tr\left[\bar\vphi_{AB}(\openone_A\otimes\sigma_B)\right]
  \right\}\\
 &=\max_{\bar\vphi_{AB}\in
    \B_*(\phi_{AB};\eps)}\left\{-\log\lambda_{\max}(\rho_B^{\bar\vphi})
  \right\}\\
  &=\max_{\bar\rho_B\in\B(\rho_B^{\phi};\eps)}S_{\min}(\bar\rho_B)\\
  &=S_{\min}^\eps(\rho_B^\phi),
\end{split}
  \end{equation}
  where in the fifth line we made use of~\reff{equi}.

  Next, we prove~\reff{eq:lemma2-stat}. By
  Lemma~\ref{lemma:accessory}, for any $P\in\P(\phi;\eps)$, the
  normalized pure state
  $|\varphi\>:=\frac{\sqrt{P}|\phi\>}{\sqrt{\Tr[P\phi]}}$ is such that
  $F\left(|\varphi\>,|\phi\>\right)\ge 1-2{\sqrt{\eps}}$, implying
  that $F^2\left(|\varphi\>,|\phi\>\right)\ge 1-4\sqrt\eps$. Let us
  define the following set, for any given bipartite pure state
  $\phi_{AB}$:
\begin{equation}\label{set1} \sA^{\eps}(\phi_{AB}) :=
  \left\{|\vphi_{AB}\rangle \in \sH_A \otimes \sH_B: |\vphi_{AB}\> =
    \frac{\sqrt{P} |\phi_{AB}\>}{\sqrt{\Tr[P\phi_{AB}]}},P\in\P(\phi_{AB};\eps)\right\}.
\end{equation}
Obviously, for $\eps'= 2\eps^{\frac 14}$, $\sA^{\eps}(\phi_{AB})\subseteq
\B_*(\phi_{AB};\eps')$, with the set $\B_*(\phi_{AB};\eps')$ being
defined by \reff{b*}. Then,

\begin{align}
  \I^{A \rightarrow B}_{0, \eps} (\phi_{AB}) =& \max_{P \in
    \P(\phi_{AB}; \eps)} \min_{\sigma_B}
  \Bigl[- \log \tr\bigl(\sqrt{P} \phi_{AB} \sqrt{P}(\openone \otimes
  \sigma_{B}) \bigr) \Bigr]\nonumber\\
=&\max_{P \in
    \P(\phi_{AB}; \eps)} \min_{\sigma_B}
  \left[- \log \tr\left(\frac{\sqrt{P} \phi_{AB} \sqrt{P}}{\Tr[P\,\phi_{AB}]}(\openone \otimes
  \sigma_{B}) \right) -\log\tr\left(P\,\phi_{AB}\right) \right]\nonumber\\
  \le& \max_{|\vphi_{AB}\rangle \in \sA^{\eps}(\phi_{AB})}
  \min_{\sigma_B}
  \Bigl[- \log \tr\bigl(\vphi_{AB} (\openone \otimes \sigma_{B})
  \bigr) \Bigr] -\log(1-\eps)\nonumber\\
  \le & \max_{|\vphi_{AB}\rangle \in \B_*(\phi_{AB}; \eps')}
  \min_{\sigma_B} \Bigl[- \log \tr\bigl(\vphi_{AB} (\openone \otimes
  \sigma_{B}) \bigr) \Bigr]-\log(1-\eps),
  \nonumber\\
  =& \max_{{\bar{\rho}}_{B} \in \B(\rho_{B}^\phi; \eps')}
  \min_{\sigma_B} \Bigl[- \log \tr\bigl({\bar{\rho}}_{B} \sigma_{B}
  \bigr) \Bigr]-\log(1-\eps),
  \nonumber\\
  =& \max_{{\bar{\rho}}_{B} \in \B(\rho_{B}^\phi; \eps')}\bigl[- \log \lambda_{\max} ({\bar{\rho}}_{B})\bigr]-\log(1-\eps) \nonumber\\
  =& S^{\eps'}_{\min} (\rho_{B}^\phi)-\log(1-\eps)\\
 =& S^{\eps'}_{\min}
  (\rho_{A}^\phi) -\log(1-\eps),
\end{align}
where $\rho_{B}^\phi := \tr_A \phi_{AB}$ and $\rho^{A}_\phi := \tr_B
\phi_{AB}$. In the above, the second inequality follows from the fact
that $\sA^{\eps}(\phi_{AB}) \subseteq \B_*(\phi_{AB};\eps')$, the
third identity follows from the fact that $ \B_*(\phi_{AB}; \eps')=
\B(\rho_{B}^\phi; \eps')$ as stated in \reff{equi}, and the last
identity holds because $\phi_{AB}$ is a pure state.
\end{proof}

\section{Distillable entanglement of a single pure state}

In order to approach the problem of quantifying the one-shot EoA of an
arbitrary bipartite mixed state, we start from the simple but
insightful case in which two distant parties, say Alice and Bob,
initially share a single copy of a {\em{pure state}}
$|\phi_{AB}\rangle$. Their aim is to distill entanglement from this
shared state (i.e., convert the state to a maximally entangled state)
using local operations and classical communication (LOCC) only.
%
For sake of generality, we consider the situation where, for any given
$\eps\ge 0$, the final state of the protocol is $\eps$-close to a
maximally entangled state, with respect to a suitable distance
measure. More precisely, we require the fidelity \reff{fidelity-aaa}
between the final state of the protocol and a maximally entangled
state to be $\ge 1 - \eps$.
\begin{definition}[$\eps$-achievable distillation rates for pure states\footnote{For the more general case of mixed states, see \cite{distil}}] 
  For any given $\eps\ge0$, a real number $R\ge 0$ is said to be an
  \emph{$\eps$-achievable rate} for one-shot entanglement distillation 
of a pure state $\phi_{AB}:=|\phi_{AB}\rangle\langle\phi_{AB}|$, if there exists an integer $M\ge 2^R$ and a maximally entangled state
  $\Psi^M_{A'B'}$ such that
\begin{equation}
F^2\left(\Lambda(\phi_{AB}),\Psi^M_{A'B'}\right)\ge 1-\eps , 
\end{equation}
for some LOCC operation $\Lambda:AB\mapsto
A'B'$. 
\end{definition}
\begin{definition}[One-shot pure-state distillable entanglement] For any given
  $\eps\ge0$, the one-shot distillabe entanglement, $E_D(\phi_{AB};\eps)$,  of a pure state $\phi_{AB}$ is the maximum of all $\eps$-achievable
  entanglement distillation rates for the state $\phi_{AB}$. 
 \end{definition}
\medskip

Bounds on the one-shot distillable entanglement of a pure state $\phi_{AB}$
are given by the following theorem.
\bigskip

\framebox[0.95\linewidth]{
\begin{minipage}{0.90\linewidth} 

\begin{theorem}\label{theo:pure}
For any bipartite pure state $\phi_{AB}$ and any $\eps\in[0,\frac 14)$,
\begin{equation}
  \label{eq:20}
  S_{\min}^\eps(\rho^\phi_A)-\Delta\le E_D(\phi_{AB};\eps)\le S_{\min}^{\eps'}(\rho^\phi_A)-\log(1-2\sqrt{\eps}),
\end{equation}
where $\rho^\phi_A:=\Tr_B\phi_{AB}$, $\eps'=2^{\frac 54}\eps^{\frac 18}$, and
$0\le \Delta \le 1$ is a number included to ensure that the lower
bound in \reff{eq:20} is the logarithm of an integer number.
\end{theorem}
\end{minipage}}
\bigskip

\begin{remark}
  The above theorem shows that, for any given $\eps\ge0$, the smoothed
  min-entropy $S_{\min}^\eps(\rho^\phi_A)$ essentially characterizes
  the one-shot distillable entanglement of the bipartite pure state
  $|\phi_{AB}\>$. In particular, for \emph{perfect} one-shot
  environment-assisted entanglement distillation, i.e. $\eps=0$,
  we obtain the identity
\begin{equation}\label{eq:e0}
E_D(\phi_{AB};0)=\log\lfloor2^{S_{\min}(\rho^\phi_A)}\rfloor.
\end{equation}
\end{remark}

\begin{remark}
  It is interesting to compare the lower bound of
  Theorem~\ref{theo:pure} with the one-shot hashing bound proved in
  Lemma~2 of~\cite{distil} for an arbitrary (possibly mixed) state.
  For pure states, using Lemma~\ref{lemma:opsm-stsm}, the latter
  yields:
  \begin{equation}\label{eq:hashing}
    E_D(\phi_{AB};\eps)\ge
    S_{\min}^{\eps/8}(\rho^\phi_A)+\log\left(\frac 1d+\frac{\eps^2}{4}\right)-\Delta,
  \end{equation}
  where $d=\dim\sH_A$. It is evident that the lower bound in
  Theorem~\ref{theo:pure} is tighter than~\reff{eq:hashing}, in
  particular because it does not have any explicit logarithmic
  dependence on the smoothing parameter $\eps$. (For example, in
  contrast to~(\ref{eq:20}), the above inequality provides a trivial
  bound in the case $\eps=0$). From the technical point of view, this
  arises as an artifact of random coding arguments used to
  derive~(\ref{eq:hashing}), whereas, for the case of pure states, we
  can directly employ Nielsen's majorization criterion.
\end{remark}

The proof of Theorem~\ref{theo:pure} can be divided into the following 
two lemmas.

\begin{lemma}
For any bipartite pure state $\phi_{AB}$ and any $\eps\ge 0$,
\begin{equation}
  \label{eq:20a}
E_D(\phi_{AB};\eps)\ge S_{\min}^\eps(\rho^\phi_A)-\Delta,
\end{equation}
where $\Delta\ge0$ is the least number such that the left hand side is
equal to the logarithm of a positive integer.
\end{lemma}

\begin{proof}
  Let us begin by considering the case $\eps=0$. In this case,
  Nielsen's majorization theorem \cite{nielsen-maj} implies that,
  using LOCC, it is possible to exactly convert any pure state
  $|\phi_{AB}\>$ to a maximally entangled state of rank equal to
  $\left\lfloor\frac 1{\lambda_{\max}}\right\rfloor$, where
  $\lambda_{\max}$ denotes the maximum eigenvalue of the reduced
  density matrix $\rho^\phi_A$. Using the definition
  (\ref{smooth_min}) of the min-entropy we then infer that
  \begin{equation}
    \label{eq:21}
    E_D(\phi_{AB};0)\ge \log\left\lfloor 2^{S_{\min}(\rho^\phi_A)}\right\rfloor.
  \end{equation}

  If we allow a finite accuracy in the conversion, a lower bound to
  the distillable entanglement can be given as follows.

  For any $|\bar\phi_{AB}\>\in\B_*(\phi_{AB};\eps)$, by Nielsen's
  theorem, there exists an LOCC map $\bar\Lambda$ such that
  \begin{equation}\label{eq:inter-fid}
    F^2\left(\bar\Lambda\left(\bar\phi_{AB}\right),\Psi_{A'B'}^{\bar M}\right)=1,
  \end{equation}
  where $\log\bar
  M:=S_{\min}\left(\rho^{\bar\phi}_A\right)$.

On the other hand, due to the monotonicity of fidelity under the
action of a completely positive trace-preserving map,
\begin{equation}
\begin{split}
  1-\eps\le1-\eps^2&\le F^2(\bar\phi_{AB},\phi_{AB})\\
  &\le F^2\left(\bar\Lambda\left(\bar\phi_{AB}\right),\bar\Lambda(\phi_{AB})\right)\\
  &=F^2\left(\Psi_{A'B'}^{\bar M},\bar\Lambda(\phi_{AB})\right).
\end{split} 
\end{equation}
This yields the bound $ E_D(\phi_{AB};\eps)\ge\log\bar M$, for any
$|\bar\phi_{AB}\>\in\B_*(\phi_{AB};\eps)$. In particular, we have that
\begin{equation}
    \label{eq:22}
    E_D(\phi_{AB};\eps)\ge \max_{\bar\phi_{AB}\in\B_*(\phi_{AB};\eps)}\log\left\lfloor 2^{S_{\min}(\rho^{\bar\phi}_A)}\right\rfloor.
    \end{equation}
    Since the two sets
    $\{\Tr_B[\bar\phi_{AB}]:\bar\phi_{AB}\in\B_*(\phi_{AB})\}$ and
    $\B(\rho^\phi_A;\eps)$ coincide~\cite{marco}, we finally arrive at
  \begin{equation}
    \label{eq:23}
    E_D(\phi_{AB};\eps)\ge \log\left\lfloor2^{S_{\min}^\eps(\rho^\phi_A)}\right\rfloor.
  \end{equation}
\end{proof}

\begin{lemma}
  For any bipartite pure state $\phi_{AB}$ and any $\eps\in[0,\frac
  14)$,
\begin{equation}
  \label{eq:20aa}
 E_D(\phi_{AB};\eps)\le S_{\min}^{\eps'}(\rho^\phi_A)-\log(1-2\sqrt{\eps}),
\end{equation}
for $\eps'=2^{\frac 54}\eps^{\frac 18}$.
\end{lemma}

\begin{proof}
  Let $r$ be the maximum of all achievable rates of entanglement
distillation for the pure state $\phi_{AB}$, i.e. $\log
  r=E_D(\phi_{AB};\eps)$. This means that there exists an LOCC
  transformation $\Lambda$ that maps $|\phi_{AB}\>$ into a state
  $\omega_{A'B'}=\Lambda(\phi_{AB})$ which is $\eps$-close to a
  maximally entangled state $|\Psi^r_{A'B'}\>$ of rank $r$, i.e.,
$F^2\left(\Lambda(\phi_{AB}),\Psi^r_{A'B'}\right) \ge 1 - \eps$. Then,
  \begin{equation}
    \label{eq:24}
\begin{split}
  E_D(\phi_{AB};\eps)&=\log r\\
  &=I^{A'\to B'}_0(\Psi^r_{A'B'})\\
  &\le \I^{A'\to B'}_{0,\eps}\left(\Lambda(\phi_{AB})\right)\\
  &\le \I^{A\to B}_{0,2\sqrt{\eps}}(\phi_{AB})\\
  &\le S_{\min}^{\eps'}(\rho^\phi_A)-\log(1-2\sqrt{\eps}),
\end{split} 
 \end{equation}
 for $\eps'=2^{\frac 54}\eps^{\frac 18}$, where the first, second and third
 inequalities follow from Lemma \ref{lemma1}, Lemma \ref{lemma2} and
 Lemma \ref{lemma:opsm-stsm}, respectively.
\end{proof}

\section{One-shot entanglement of assistance}

As stated in the introduction, the definition of the EoA
arises naturally when considering the task in which Alice and Bob
distill entanglement from an initial mixed bipartite state $\rho_{AB}$
which they share, when a third party (say Charlie), who holds the
purification of the state, assists them, by doing local operations on
his share and communicating classical bits to Alice and Bob.

In order to express these ideas in a mathematically sound form, we
start by noticing that any strategy that Charlie may employ can be
described as the measurement of a positive operator-valued measure
(POVM) $\{P^i_C\}_i$, followed by the communication, to both Alice and
Bob, of the resulting classical outcome $i$. Since the state shared
between Alice, Bob, and Charlie is pure, say $|\Psi^\rho_{ABC}\>$,
Charlie's POVM's are in one-to-one correspondence with decompositions
of $\rho_{AB}$ into ensembles $\{p_i,\rho^i_{AB}\}_i$, via the
relation $p_i\rho^i_{AB}:=\Tr_C[\Psi^\rho_{ABC}\ (\openone_{AB}\otimes
P^i_C)]$. The fact that Charlie announces which outcome he got, means
that Alice and Bob can apply a different LOCC map for each value of
$i$.

An important point to stress now is that, in general, the distillation
process is allowed to be approximate. This is needed, in particular,
if one later wants to recover, from the one-shot setting, the usual
asymptotic scenario, where errors are required to vanish
asymptotically but are finite otherwise. In the classically-assisted
case we are studying here, since the index $i$ is visible to Alice and
Bob, they can apply a different LOCC map $\Lambda_i$ for each state
$\rho^i_{AB}$. We can hence choose to evaluate the distillation
accuracy according to a worst-case or an average criterion. Here we
choose the average fidelity as a measure of the ``expected''
accuracy. This leads us to define the maximum amount of entanglement
that can be distilled in the assisted case, namely, the \emph{one-shot
  entanglement of assistance}, as,
\begin{equation}\label{eq:eoa-def}
\begin{split}
 & D_A(\rho_{AB};\eps)\\
&:=\max_{\{P^i_C\}_i}\max_{M\in\mathbb{N}}\left\{
\log M: \max_{\{\Lambda^i_{AB}\}_i}
    F^2\left(\sum_ip_i\Lambda^i(\rho^i_{AB}),\Psi^M_{A'B'}\right)\ge 1-\eps
\right\},
\end{split}
\end{equation}
where each $\Lambda^i$ is an LOCC map from $AB$ to $A'B'$.

As proved in Appendix~\ref{app:A}, the maximization over Charlie's
measurement in the above definition can always be restricted, without
loss of generality, to rank-one POVM's. Since rank-one POVM's at
Charlie's side are in one-to-one correspondence with pure state
ensemble decompositions of $\rho_{AB}$, we can equivalently write
\begin{equation}\label{eq:eoa-def2}
\begin{split}
  &D_A(\rho_{AB};\eps)\\
=&\max_{{\{p_i,\phi^i_{AB}\}_i}\atop{\sum_ip_i\phi^i_{AB}=\rho_{AB}}}\max_{M\in\mathbb{N}}\left\{
\log M: \max_{\{\Lambda^i_{AB}\}_i}
    F^2\left(\sum_ip_i\Lambda^i(\phi^i_{AB}),\Psi^M_{A'B'}\right)\ge 1-\eps
\right\}.
\end{split}
\end{equation}

In order to quantify $D_A(\rho_{AB};\eps)$ then, it is sufficient to
quantify the maximum expected amount of entanglement that can be
distilled, in average, from any given ensemble of pure bipartite
states. This is the aim of the following section.

\section{Distillable entanglement of an ensemble of pure states}

Given an ensemble $\eE=\{p_i,\phi^i_{AB}\}$ of pure states, we
define, for any given $\eps\ge0$ the one-shot distillable entanglement
of $\eE$ as
\begin{equation}\label{eq:ede-def}
  E_D(\eE;\eps):=\max_{M\in\mathbb{N}}\left\{\log M:\max_{\{\Lambda^i_{AB}\}_i}
    F^2\left(\sum_ip_i\Lambda^i(\phi^i_{AB}),\Psi^M_{A'B'}\right)\ge 1-\eps\right\},
\end{equation}
where each $\Lambda^i$ is an LOCC map from $AB$ to $A'B'$. According
with equation~(\ref{eq:eoa-def2}), the one-shot entanglement of
assistance $E_A$ of a given mixed state $\rho_{AB}$ is given by
\begin{equation}\label{ea}
 D_A(\rho_{AB};\eps)=\max_{\eE}E_D(\eE;\eps),
\end{equation}
where the maximum is over all possible pure state ensemble
decompositions $\eE$ of $\rho_{AB}$.

For any given ensemble $\eE=\{p_i, \phi^i_{AB}\}$ of pure states, we
define the quantity
\begin{equation}\label{eff}
F_{\min}(\eE) := \min_i
S_{\min}(\rho^{\phi^i}_A),
\end{equation}
where $\rho^{\phi^i}_A:= \tr_B {\phi^i_{AB}}$. This quantity can be
intuitively interpreted as a conservative estimate of the amount of
entanglement present in the ensemble $\eE$. Further, for any such
ensemble, and any given $\eps \ge 0$, let us define the set
\begin{equation}\label{set11}
  \cS_{\lle}(\eE;\eps):=\left\{\bar\eE=\left\{\bar\vphi^i_{AB}\right\}_i:\Tr\bar\vphi^i_{AB}\le 1, {\sum_ip_iF(\bar\vphi^i_{AB},\phi^i_{AB})\ge 1- {{\eps}}}\right\},
\end{equation}
and let $\cS_{\=}(\eE;\eps)$ denote the set obtained from 
$\cS_{\lle}(\eE;\eps)$ by restricting the
pure states $\bar\vphi^i_{AB}$ to be normalized.\bigskip

\framebox[0.95\linewidth]{
\begin{minipage}{0.90\linewidth} 
\begin{theorem}\label{thm_2}
For any given ensemble $\eE=\{p_i, \phi^i_{AB}\}$ of pure states, and any $\eps\ge 0$,
\begin{equation}\label{stat}
 \max_{\bar\eE\in \cS_{=}(\eE ;\eps')}F_{\min}(\bar\eE) - \Delta
\ \le\   E_D(\eE;\eps)\ \le\ 
\max_{\bar\eE\in \cS_{\lle}(\eE; \eps'')}F_{\min}(\bar\eE),
\end{equation}
where $\eps'= \eps/2$, $\eps'':= \sqrt{2 \sqrt{\eps}}$, and $0\le
\Delta \le 1$ is a number which is included to ensure that the lower
bound in \reff{stat} is the logarithm of an integer number.
\end{theorem}
\end{minipage}
}\bigskip

As a note, we explicitly remark that Theorem~\ref{thm_2} gives the
following characterization of the one-shot entanglement of assistance
for $\eps=0$:
\begin{equation}
D_A(\rho_{AB};0)=\max_{\eE}F_{\min}(\eE),
\end{equation}
where the maximum is over all possible pure state ensemble
decompositions $\eE$ of $\rho_{AB}$.

\noindent The proof of Theorem~\ref{thm_2} is divided into the
following two lemmas.
\begin{lemma}[Direct part]
  For any pure state ensemble $\eE=\{p_i,\phi^i_{AB}\}$ and any
  $\eps\ge 0$,
\begin{equation}
  E_D(\eE;\eps)\ge\max_{\bar\eE\in \cS_{=}(\eE ;\eps')}F_{\min}(\bar\eE)-\Delta,
\end{equation}
where $\Delta$ is the minimum number in $[0,1]$ such that the right
hand side is equal to the logarithm of an integer number $M\ge 1$. 
\end{lemma}
\begin{proof}
  From Theorem~\ref{theo:pure}, we know that, given the pure bipartite
  state $\phi^i_{AB}$, Alice and Bob can distill $\log
  \left\lfloor2^{S_{\min}\left(\rho^{\phi^i}_A\right)}\right\rfloor$
  ebits with zero error. Hence, given the ensemble
  $\eE=\{p_i,\phi^i_{AB}\}$, Alice and Bob can distill, {\em{without
      error}}, at least $\min_i \log
  \left\lfloor2^{S_{\min}\left(\rho^{\phi^i}_A\right)} \right\rfloor$
  ebits. For any pure state ensemble $\eE$, let us then introduce the
  quantity $M(\eE):=\min_i
  \left\lfloor2^{S_{\min}\left(\rho^{\phi^i}_A\right)} \right\rfloor$.

  If a finite accuracy $\eps> 0$ is allowed, then it is possible to
  give a lower bound on the one-shot distillable entanglement
  $E_D(\eE;\eps)$ as follows. Let us consider the set of ensembles of
  normalized pure states of the form $\bar\eE=\{p_i,
  \bar\vphi^i_{AB}\}$, such that
  $\sum_ip_iF(\phi^i_{AB},\bar\vphi^i_{AB})\ge 1-\eps$. Then, for any
  ensemble $\bar\eE$ in such a set, there exist LOCC maps
  $\Lambda^i:AB\rightarrow A'B'$ such that
  \begin{equation}\label{eq:exact-ens-dist}
    F\left(\sum_ip_i\Lambda^i(\bar\vphi^i_{AB}),\Psi^{M(\bar\eE)}_{AB}\right)=1,
  \end{equation}
  where $\Psi^{M(\bar\eE)}_{A'B'}$ denotes a maximally entangled state
  of rank $M(\bar\eE)$. Equivalently,
  $\Lambda^i(\bar\vphi^i_{AB})=\Psi^{M(\bar\eE)}_{AB}$, for all
  $i$. Then,
  \begin{equation}
\begin{split}
   1-\eps&\le \sum_ip_iF(\phi^i_{AB},\bar\vphi^i_{AB})\\
&\le
\sum_ip_iF\left(\Lambda^i(\phi^i_{AB}),\Lambda^i(\bar\vphi^i_{AB})\right)\\
&\le
F\left(\sum_ip_i\Lambda^i(\phi^i_{AB}),\sum_ip_i\Lambda^i(\bar\vphi^i_{AB})\right)\\
&=F\left(\sum_ip_i\Lambda^i(\phi^i_{AB}),\Psi^{M(\bar\eE)}_{AB}\right),
\end{split}
  \end{equation}
  where the second line follows from the monotonicity of fidelity
  under completely positive trace-preserving (CPTP) maps, the third
  line follows from the concavity of the fidelity, and the last
  identity follows from~(\ref{eq:exact-ens-dist}). Hence, we conclude
  that there exist LOCC maps $\Lambda^i$ for which
\begin{equation}
 F^2\left(\sum_ip_i\Lambda^i(\phi^i_{AB}),\Psi^{M(\bar\eE)}_{A'B'}\right)\ge 1-2\eps,
\end{equation}
that is,
\begin{equation}
 E_D(\eE;{2\eps})\ge\log M(\bar\eE),
\end{equation}
{\em{for any}} $\bar\eE$ in the set introduced above.
By maximizing $M(\bar\eE)$ over
all such ensembles and comparing the result with the
definition in~(\ref{eff}),
we obtain the statement of the lemma.
\end{proof}

\begin{lemma}[Converse part]\label{lemma:eoe-conv}
For any pure state ensemble $\eE=\{p_i,\phi^i_{AB}\}$ and any $\eps\ge 0$,
\begin{equation}
E_D(\eE;\eps)\le \max_{\bar\eE\in \cS_{\lle}(\eE; \eps')}F_{\min}(\bar\eE),
\end{equation}
where $\eps'= \sqrt{2\sqrt{\eps}}$.
%
\end{lemma}

\begin{proof}
  Let $r$ be a positive integer such that $E_D(\eE;\eps)=\log
  r$. According to~\reff{eq:ede-def}, this means that there exist LOCC
  maps $\Lambda^i:AB\to A'B'$ such that
  \begin{equation}\label{eq:av-fid-cond}
\Tr\left[
 \sum_ip_i\Lambda^i(\phi^i_{AB})\ 
\Psi^{r}_{A'B'}
\right]\ge 1-\eps.
  \end{equation}
  Since the maps $\Lambda^i$ act on pure states, without loss of
  generality we can assume them to be of the Lo-Popescu form
  \reff{form}.

Further, equation~(\ref{eq:av-fid-cond}) above, in particular, informs us that
\begin{equation}\label{65}
  \Psi^r_{A'B'}\in\P\left(\sum_ip_i\Lambda^i(\phi^i_{AB});\eps\right).
\end{equation}
This fact in turns implies that
\begin{equation}\label{eq:tocontt}
\begin{split}
  E_D(\eE;\eps)&=\log r\\
  &=I^{A'\to B'}_{0}\left(\Psi^r_{A'B'}\right)\\
  &\equiv\min_{\sigma_{B'}}\left\{-\log\Tr\left[\Psi^r_{A'B'}\
      (\openone_{A'}\otimes\sigma_{B'})\right]\right\}\\
  &\le -\log\Tr\left[\Psi^r_{A'B'}\
    (\openone_{A'}\otimes\tilde\sigma_{B'})\right]\\
  &\le-\log\Tr\left[\left(\Psi^r_{A'B'}\
      \Pi_{\sum_ip_i\Lambda^i(\phi^i_{AB})}\
      \Psi^r_{A'B'}\right)\
    (\openone_{A'}\otimes\tilde\sigma_{B'})\right],
\end{split}
\end{equation}
for any state $\tilde\sigma_{B'}$. To obtain the last inequality, 
we simply used
the fact that $\Psi^r_{A'B'}\ge \Psi^r_{A'B'}\Pi \Psi^r_{A'B'}$, for
any $0\le\Pi\le\openone$. We then choose $\tilde\sigma_{B'}$ so that
\begin{equation}\label{68}
\begin{split}
  &-\log\Tr\left[\left(\Psi^r_{A'B'}\
      \Pi_{\sum_ip_i\Lambda^i(\phi^i_{AB})}\
      \Psi^r_{A'B'}\right)\
    (\openone_{A'}\otimes\tilde\sigma_{B'})\right]\\
  =\min_{\sigma_{B'}}&\left\{-\log\Tr\left[\left(\Psi^r_{A'B'}\
        \Pi_{\sum_ip_i\Lambda^i(\phi^i_{AB})}\
        \Psi^r_{A'B'}\right)\
      (\openone_{A'}\otimes\sigma_{B'})\right]\right\}.
\end{split}
\end{equation}
From \reff{65}, \reff{eq:tocontt} and \reff{68} we infer that
\begin{equation}
  E_D(\eE;\eps)\le
\I^{A'\to
  B'}_{0,\eps}\left(\sum_ip_i\Lambda_i(\phi^i_{AB})\right).
\end{equation}

Let us now introduce an auxiliary system $Z$ and an orthonormal basis
for it $\{|i_Z\>\}$ that keeps track of the classical outcome $i$
labeling the states in $\eE$. Let us denote by $\pi^i_Z$ the projector
$|i\>\<i|_Z$. By further introducing the states
$\omega_{A'B'}:=\sum_ip_i\Lambda_i(\phi^i_{AB})$ and
$\omega_{A'B'Z}:=\sum_ip_i\Lambda_i(\phi^i_{AB})\otimes\pi^i_Z$, so
that $\omega_{A'B'}=\Tr_Z \omega_{A'B'Z}$, we have
\begin{equation}
\begin{split}
  E_D(\eE;\eps) &\le\I^{A'\to  B'}_{0,\eps}(\omega_{A'B'})\\
&\equiv\max_{P\in\P(\omega_{A'B'};\eps)}\min_{\sigma_{B'}}\left\{
  -\log\Tr\left[ \sqrt{P}\Pi_{\omega_{A'B'}}\sqrt{P}\
    (\openone_{A'}\otimes\sigma_{B'})\right]\right\}\\
&=\min_{\sigma_{B'}}\left\{
  -\log\Tr\left[ \sqrt{P_0}\Pi_{\omega_{A'B'}}\sqrt{P_0}\
    (\openone_{A'}\otimes\sigma_{B'})\right]\right\}\\
&\le  -\log\Tr\left[ \sqrt{P_0}\Pi_{\omega_{A'B'}}\sqrt{P_0}\
    (\openone_{A'}\otimes\bar\nu_{B'})\right],
\end{split} 
\end{equation}
where the operator $P_0$ in the third line is the one achieving the
maximum, and $\bar\nu_{B'}$ in the fourth line is any state in
$\states(\sH_{B'})$. In particular, since
$\Pi_{\omega_{A'B'}}\otimes\openone_Z \ge\Pi_{\omega_{A'B'Z}}$, we have that
\begin{equation}
\begin{split}
  E_D(\eE;\eps)&\le -\log\Tr\left[
    \sqrt{P_0}\Pi_{\omega_{A'B'}}\sqrt{P_0}\
    (\openone_{A'}\otimes\bar\nu_{B'})\right]\\
  &= -\log\Tr\left[
    \sqrt{P_0\otimes\openone_Z}(\Pi_{\omega_{A'B'}}\otimes\openone_Z)\sqrt{P_0\otimes\openone_Z}\
    (\openone_{A'}\otimes\bar\nu_{B'Z})\right]\\
&\le  -\log\Tr\left[
    \sqrt{P_0\otimes\openone_Z}\Pi_{\omega_{A'B'Z}}\sqrt{P_0\otimes\openone_Z}\
    (\openone_{A'}\otimes\bar\nu_{B'Z})\right],
\end{split}
\end{equation}
for any state $\bar\nu_{B'Z}$.

Let us then choose $\bar\nu_{B'Z}$ to be the state such that
\begin{equation}
\begin{split}
 &-\log\Tr\left[
    \sqrt{P_0\otimes\openone_Z}\Pi_{\omega_{A'B'Z}}\sqrt{P_0\otimes\openone_Z}\
    (\openone_{A'}\otimes\bar\nu_{B'Z})\right]\\
=\min_{\nu_{B'Z}}&\left\{-\log\Tr\left[
    \sqrt{P_0\otimes\openone_Z}\Pi_{\omega_{A'B'Z}}\sqrt{P_0\otimes\openone_Z}\
    (\openone_{A'}\otimes\nu_{B'Z})\right]\right\}.
\end{split}
\end{equation}
Moreover, note that
$(P_0\otimes\openone_Z)\in\P(\omega_{A'B'Z};\eps)$, since
$P_0\in\P(\omega_{A'B'};\eps)$.
In fact, the operator $(P_0 \otimes \openone_Z)$ also belongs
to the following set of quantum-classical (q-c) operators:
\begin{equation}\label{qc}
\begin{split}
  &\P_{\textrm{qc}}(\omega_{A'B'Z};\eps):=\\
&\left\{\left.P_{A'B'Z}=\sum_iP^i_{A'B'}\otimes\pi^i_Z\right|
  0\le P^i_{A'B'} \le \openone_{A'B'},\, \Tr\bigl(
  P_{A'B'Z}\,\omega_{A'B'Z}\bigr)\ge1-\eps\right\}.
\end{split}
\end{equation}

Hence, we can write
\begin{equation}\label{eq:qwerty}
  E_D(\eE;\eps)\le\max_{Q\in\P_{\textrm{qc}}(\omega_{A'B'Z};\eps)}\min_{\nu_{B'Z}}\left\{-\log\Tr\left[
      \sqrt{Q}\Pi_{\omega_{A'B'Z}}\sqrt{Q}\
      (\openone_{A'}\otimes\nu_{B'Z})\right]\right\}
\end{equation}

Let the Kraus representations of the CPTP maps $\Lambda_i:AB\mapsto
A'B'$ satisfying~\reff{eq:av-fid-cond} be written as
$\Lambda_i(\rho)=\sum_{\mu_i}V_{\mu_i}\rho V_{\mu_i}^\dag$, so that
$\sum_{\mu_i}V_{\mu_i}^\dag V_{\mu_i}=\openone_{AB}$ for all
$i$. Using these, we construct a CPTP map $\M:ABZ\to A'B'Z$ as
\begin{equation}
  \label{eq:11}
  \M(\rho_{ABZ}):=\sum_i\sum_{\mu_i}\left(V_{\mu_i}\otimes\pi^i_Z\right)\rho_{ABZ}\left(V_{\mu_i}\otimes\pi^i_Z\right)^\dag.
\end{equation}
In terms of the map $\M$ so constructed,
\begin{equation}
 \omega_{A'B'Z}=\M\left(\sum_ip_i\phi^i_{AB}\otimes\pi^i_Z\right).
\end{equation}
Defining the quantum-classical (q-c) state
$\sigma_{ABZ}:=\sum_ip_i\phi^i_{AB}\otimes\pi^i_Z$, we have,
continuing from~(\ref{eq:qwerty}),
\begin{align}
E_D(\eE;\eps)&\le
\max_{Q\in\P_{\textrm{qc}}(\M(\sigma_{ABZ});\eps)}\min_{\nu_{B'Z}}\left\{-\log\Tr\left[\sqrt{Q}\Pi_{\M(\sigma_{ABZ})}\sqrt{Q}\
    \left(\openone_{A'}\otimes\nu_{B'Z}\right)\right]\right\}\nonumber\\
&\equiv
\min_{\nu_{B'Z}}\left\{-\log\Tr\left[\sqrt{Q_0}\Pi_{\M(\sigma_{ABZ})}\sqrt{Q_0}\
    \left(\openone_{A'}\otimes\nu_{B'Z}\right)\right]\right\},
\end{align}
where $Q_0\in\P_{\textrm{qc}}(\M(\sigma_{ABZ});\eps)$ is the q-c
operator achieving the maximum in the second line. This implies that
\begin{equation}
  E_D(\eE;\eps)\le -\log\Tr\left[\sqrt{Q_0}\Pi_{\M(\sigma_{ABZ})}\sqrt{Q_0}\ 
    \left(\openone_{A'}\otimes\nu_{B'Z}\right)\right],
\end{equation}
for any state $\nu_{B'Z}$.

Due to the fact that the maps $\Lambda_i$ are in the Lo-Popescu
form \reff{form}, it follows that the map $\M$ (obtained from the
$\Lambda_i$'s) is also in the Lo-Popescu form. The identity \reff{lopop}
then implies that
\begin{equation}\label{eq:quasi-fin}
   E_D(\eE;\eps)\le -\log\Tr\left[\sqrt{Q_0}\Pi_{\M(\sigma_{ABZ})}\sqrt{Q_0}\ 
    \ \M(\openone_{A}\otimes\tilde\nu_{BZ})\right],
\end{equation}
for any state $\tilde\nu_{BZ}$. By using the dual map $\M^*$,
\begin{equation}
    E_D(\eE;\eps)\le -\log\Tr\left[\M^*\left(\sqrt{Q_0}\Pi_{\M(\sigma_{ABZ})}\sqrt{Q_0}\right)\ 
    \ (\openone_{A}\otimes\tilde\nu_{BZ})\right],
\end{equation}
for any state $\tilde\nu_{BZ}$. By denoting the operator
$\M^*\left(\sqrt{Q_0}\Pi_{\M(\sigma_{ABZ})}\sqrt{Q_0}\right)$ as $\tilde
Q_{ABZ}$, we have, for any state $\tilde\nu_{BZ}$,
\begin{equation}\label{eq:to_cont}
     E_D(\eE;\eps)\le -\log\Tr\left[\sqrt{\tilde
         Q_{ABZ}}\Pi_{\sigma_{ABZ}}\sqrt{\tilde Q_{ABZ}}\ (\openone_{A}\otimes\tilde\nu_{BZ})\right],
\end{equation}
since $\tilde Q_{ABZ}\ge \sqrt{\tilde
  Q_{ABZ}}\Pi_{\sigma_{ABZ}}\sqrt{\tilde Q_{ABZ}}$. Let us also
choose $\tilde\nu_{BZ}$ so that
\begin{equation}
 \begin{split}
&-\log\Tr\left[\sqrt{\tilde
         Q_{ABZ}}\Pi_{\sigma_{ABZ}}\sqrt{\tilde Q_{ABZ}}
    \ (\openone_{A}\otimes\tilde\nu_{BZ})\right]\\
=\min_{\nu_{BZ}}&\left\{-\log\Tr\left[\sqrt{\tilde
         Q_{ABZ}}\Pi_{\sigma_{ABZ}}\sqrt{\tilde Q_{ABZ}} 
    \ (\openone_{A}\otimes\nu_{BZ})\right]\right\}.
\end{split}
\end{equation}

Using the particular form~(\ref{eq:11}) of $\M$, and the facts that
$\sigma_{ABZ}$ is a q-c state and $Q_0 \in
\P_{\textrm{qc}}(\M(\sigma_{ABZ});\eps)$, we can prove that the operator
$\tilde Q_{ABZ} \in \P_{\textrm{qc}}(\sigma_{ABZ};2\sqrt{\eps})$,
using arguments similar to those leading to \reff{33}.

Hence, continuing from equation~\reff{eq:to_cont}, we can write
\begin{align}
  E_D(\eE;\eps)&\le  \min_{\nu_{BZ}}\left\{-\log\Tr\left[\sqrt{\tilde
         Q_{ABZ}}\Pi_{\sigma_{ABZ}}\sqrt{\tilde Q_{ABZ}} 
    \ (\openone_{A}\otimes\nu_{BZ})\right]\right\}\nonumber\\
&\le \max_{P\in \P_{\textrm{qc}}(\sigma_{ABZ};2\sqrt{\eps})}\min_{\nu_{BZ}} \left\{-\log\Tr\left[\sqrt{P}\Pi_{\sigma_{ABZ}}\sqrt{P}\ 
    \ (\openone_{A}\otimes\nu_{BZ})\right]\right\}.\label{81}
\end{align}

Let $\eps' := 2\sqrt{\eps}$. Then, for any
$P=\sum_iP^i_{AB}\otimes\pi^i_Z$ in
$\P_{\textrm{qc}}(\sigma_{ABZ};\eps')$, let us define
$|\vphi_{AB}^i\rangle := \sqrt{P^i_{AB}}|\phi_{AB}^i\rangle$. As a
consequence of Lemma~\ref{fid3}, we have that
$\sum_ip_iF(\vphi_{AB}^i, \phi^i_{AB})\ge 1-\sqrt{\eps'}$, so that
\begin{align} E_D(\eE;\eps)
&\le  \max_{P\in
  \P_{\textrm{qc}}(\sigma_{ABZ};{\eps'})}\min_{\nu_{BZ}}
\left\{-\log\Tr\left[\sqrt{P}\Pi_{\sigma_{ABZ}}\sqrt{P}\
    \ (\openone_{A}\otimes\nu_{BZ})\right]\right\}\nonumber\\
&\le 
\max_{\bar\eE\in \cS_{\lle}(\eE; \sqrt{\eps'})}
\min_{\nu_{BZ}} \left\{-\log\Tr\left[\bigl(\sum_i\bar\vphi^i_{AB}\otimes
    \pi^i_{Z}\bigr)\
    \ (\openone_{A}\otimes\nu_{BZ})\right]\right\}\nonumber\\
&=\max_{\bar\eE\in \cS_{\lle}(\eE; \sqrt{\eps'})}
\min_i \min_{\nu_{B}} \left\{-\log\Tr\left[\rho_B^{\bar\vphi^i}\nu_{B}\right]\right\}\nonumber\\
&=
\max_{\bar\eE\in \cS_{\lle}(\eE; \sqrt{\eps'})}
\min_i \bigl[- \log \lambda_{\max}(\rho_B^{\bar\vphi^i})\bigr],
\nonumber\\
&=
\max_{\bar\eE\in \cS_{\lle}(\eE; \sqrt{\eps'})}\min_i
S_{\min}(\rho_A^{\bar\vphi^i}),
\end{align}
where we used the fact that $\lambda_{\max}(\rho_B^{\bar\vphi^i})=
\lambda_{\max}(\rho_A^{\bar\vphi^i})=S_{\min}(\rho_A^{\bar\vphi^i})$,
since $\bar\vphi^i_{AB}$ is a pure state.
\end{proof}

\section{Asymptotic entanglement of assistance}

Consider the situation in which three parties, Alice, Bob and Charlie
jointly possess multiple (say $n$) copies of a tripartite pure state
$|\Psi_{ABC}\rangle$. Alice and Bob, considered in isolation,
therefore possess $n$ copies of the state $\rho_{AB}:= \tr_C
\Psi_{ABC}$, i.e., they share the state $\rho_{AB}^{\otimes n}$. We
refer to this situation as the ``i.i.d. scenario'', in analogy with
the classical case of independent and identically distributed (i.i.d.)
random variables. We define the asymptotic entanglement of assistance
of a state $\rho_{AB}$ as
\begin{equation}\label{eainfty}
D_A^\infty(\rho_{AB}):= \lim_{\eps \rightarrow 0} \lim_{n \rightarrow \infty} \frac{1}{n} E_A( \rho_{AB}^{\otimes n} ; \eps),
\end{equation}
where for any $\eps \ge 0$, $D_A( \rho_{AB}^{\otimes n} ; \eps)$
denotes the one-shot entanglement of assistance of the state
$\rho_{AB}^{\otimes n}$, defined in~\reff{eq:eoa-def} and quantified
in~\reff{ea} and~\reff{stat}.

The notation $E_A^\infty(\rho_{AB})$ was used in
Ref.~\cite{assistance} to denote the \emph{regularized} EoA, formally
defined as $\lim_{n\to\infty}\frac 1nE_A(\rho_{AB}^{\otimes n})$
from~\reff{eq:19}. The aim of this section is to show that the two
quantities coincide. This provides an alternative proof of the
operational interpretation of the regularized EoA given
in~\cite{assistance}.

The main result of this section is the following theorem:\bigskip

\framebox[0.95\linewidth]{
\begin{minipage}{0.90\linewidth}
\begin{theorem}\label{thm_4}
For any bipartite state $\rho_{AB}$
\begin{equation}\label{asymp}
D_A^\infty(\rho_{AB}):=\lim_{\eps \rightarrow 0} \lim_{n \rightarrow \infty} \frac{1}{n} D_A( \rho_{AB}^{\otimes n} ; \eps)= \lim_{n \rightarrow \infty} \frac{1}{n} E_A(\rho_{AB}^{\otimes n}),
\end{equation}
where for any state $\omega_{AB}$, 
\begin{equation}\label{eadef}
  E_A(\omega_{AB}):= \max_{\{p_i, |\vphi^i_{AB}\rangle\}\atop{\omega_{AB} =
      \sum_i p_i \vphi^i_{AB}}} \sum_i p_i S(\rho^{\vphi^i}_A),
\end{equation}
denotes its entanglement of assistance, with $\rho^{\vphi^i}_A=
\tr_B[\vphi^i_{AB}]$.
\end{theorem}
\end{minipage}
}\bigskip

In order to prove this, we first need to introduce a few more
definitions. Let $\sigma_{ABZ}$ be a quantum-classical (qc) state,
i.e.
\begin{equation}
  \sigma_{ABZ}=\sum_ip_i\sigma^i_{AB}\otimes\pi^i_Z,
\end{equation}
for some probabilities $p_i\ge 0$, $\sum_ip_i=1$, some normalized
states $\sigma^i_{AB}\in\states(\sH_A\otimes\sH_B)$, and some
orthogonal rank-one projectors $\pi^i_Z=|i\>\<i|_Z$ (that we fix here
once and for all). As it has been done already in~\reff{qc}, along the
proof of Lemma~\ref{lemma:eoe-conv}, we define the sets
\begin{equation}\label{pqc2}
  \P_{\textrm{qc}}(\sigma_{ABZ};\eps):=\left\{P_{ABZ}=\sum_iP^i_{AB}\otimes\pi^i_Z\left|
\begin{split}
  &0\le P^i_{AB} \le \openone_{AB},\\
&\Tr[
  P\sigma]\ge1-\eps
\end{split}
\right.\right\},
\end{equation}
and
\begin{equation}\label{bqc2}
\begin{split}
&\B_{\textrm{qc}}(\sigma_{ABZ};\eps):=\\
&\left\{\bar\omega_{ABZ}=\sum_ip_i\bar\vphi^i_{AB}\otimes\pi^i_Z\left|
\begin{split}
  &\N{\bar\vphi^i_{AB}}_1=\N{\bar\vphi^i_{AB}}_\infty=1,\\
  &F(\bar\omega,\sigma)=\sum_ip_iF(\bar\vphi^i,\sigma^i)\ge1-\eps
\end{split}
\right.\right\} .
\end{split}
\end{equation}
The sets defined above are analogous to those introduced
in~\reff{ball} and~\reff{P-ball}, with the difference that the
quantum-classical structure of the argument $\sigma_{ABZ}$ is here
maintained.

For technical reasons that will be apparent in the proofs, we also
need to introduce an additional smoothed zero-coherent information,
besides those in~\reff{eq:i} and~\reff{eq:itilda}, defined as, for any
qc state $\sigma_{ABZ}$ and any $\eps\ge 0$,
\begin{equation}\label{eq2}
  I^{A\leadsto BZ}_{0,\eps}(\sigma_{ABZ}):=\max_{\bar\sigma_{ABZ}\in \B_{\textrm{qc}}
    (\sigma_{ABZ};\eps)}\min_{\nu_{BZ}\in\states(\sH_B\otimes \sH_Z)}
  S_0(\bar\sigma_{ABZ}\|\openone_A\otimes\nu_{BZ}).
\end{equation}

We then proceed by proving the following lemma, which is nothing but a
convenient reformulation of Theorem~\ref{thm_2}:

\begin{lemma}\label{thm_3}
For any bipartite state $\rho_{AB}$ and any $\eps\ge 0$,
\begin{equation}\label{88}
\max_{\eE}I^{A\leadsto BZ}_{0,\eps/2}(\sigma_{ABZ}^\eE) - \Delta \le D_A(\rho_{AB};\eps)\le 
\max_{\eE} \I^{A\to BZ}_{0,2\sqrt{\eps}}(\sigma_{ABZ}^\eE),
\end{equation}
where the maxima are taken over all possible pure state ensembles
$\eE=\{p_i, \phi^i_{AB}\}$ such that $\rho_{AB} = \sum_i p_i
\phi^i_{AB}$, and for a given ensemble $\eE=\{p_i, \phi^i_{AB}\}$,
$\sigma_{ABZ}^\eE = \sum_i p_i \phi^i_{AB} \otimes \pi^i_Z.$ In the
above, the real number $0\le \Delta \le 1$ is included to ensure that
the lower bound is equal to the logarithm of a positive integer.
\end{lemma}

For the sake of clarity, we divide the proof of the Lemma above into
two separate lemmas. The first is the following:

\begin{lemma}\label{cor1}
  For any given ensemble $\eE=\{p_i, \phi^i_{AB}\}$ of pure states,
  and any $\eps\ge 0$,
\begin{equation}
  E_D(\eE;\eps)\le \I^{A\to
    BZ}_{0,2\sqrt{\eps}}(\sigma_{ABZ}^\eE),
\end{equation}
where $\sigma_{ABZ}^\eE:= \sum_i p_i \phi^i_{AB} \otimes \pi^i_Z$, and
$\I^{A\to BZ}_{0,2\sqrt{\eps}}(\sigma_{ABZ}^\eE)$ is defined
in~\reff{eq:itilda}.
\end{lemma}

\begin{proof}
  The equation number~(\ref{81}) in the proof of Theorem~\ref{thm_2},
  that is,
\begin{equation}
  E_D(\eE;\eps)
\le \max_{P\in \P_{\textrm{qc}}(\sigma_{ABZ};2\sqrt{\eps})}\min_{\nu_{BZ}} \left\{-\log\Tr\left[\sqrt{P}\Pi_{\sigma_{ABZ}}\sqrt{P}\ 
    \ (\openone_{A}\otimes\nu_{BZ})\right]\right\}
\end{equation}
already proves the statement, since
$\P_{\textrm{qc}}(\sigma_{ABZ};2\sqrt{\eps})\subset
\P(\sigma_{ABZ};2\sqrt{\eps})$.
\end{proof}

\begin{lemma}\label{cor2}
  For any given ensemble $\eE=\{p_i, \phi^i_{AB}\}$ of pure states,
  and any $\eps\ge 0$,
\begin{equation}
E_D(\eE;\eps)\ge I^{A\leadsto
    BZ}_{0,\eps/2}(\sigma_{ABZ}^\eE).
\end{equation}
where $\sigma_{ABZ}^\eE:= \sum_i p_i \phi^i_{AB} \otimes \pi^i_Z$ and
$I^{A\leadsto BZ}_{0,\eps/2}(\sigma_{ABZ}^\eE)$ is defined
in~\reff{eq2}.
\end{lemma}

\begin{proof}
The statement is a direct consequence of the lower bound in Theorem
\ref{thm_2}. This can be shown as follows:
\begin{align}
I^{A\leadsto BZ}_{0,\eps/2}(\sigma_{ABZ}^\eE) :&= 
\max_{\bar{\sigma}_{ABZ}\in \B_{\textrm{qc}}({\sigma}_{ABZ};\eps/2)} \min_{\nu_{BZ}} 
\left\{-\log \tr\left[\Pi_{\bar{\sigma}_{ABZ}}\ (\openone_A \otimes \nu_{BZ})\right] \right\}\nonumber\\
& = \max_{\{\bar\vphi^i_{AB}\}_i:\Tr\bar\vphi^i_{AB}=1\atop{\sum_ip_iF(\bar\vphi^i_{AB},\phi^i_{AB})\ge1-\eps/2}}\min_i \min_{\nu_B} 
\left\{ - \log\tr\left[\rho^{\bar\vphi^i}_{B}\ \nu_B\right]\right\}\nonumber\\
&= \max_{\{\bar\vphi^i_{AB}\}_i:\Tr\bar\vphi^i_{AB}=1\atop{\sum_ip_iF(\bar\vphi^i_{AB},\phi^i_{AB})\ge1-\eps/2}}\min_i \left\{ - \log \lambda_{\max}
\left(\rho^{\bar\vphi^i}_B\right)\right\}\nonumber\\
& =  \max_{\{\bar\vphi^i_{AB}\}_i:\Tr\bar\vphi^i_{AB}=1\atop{\sum_ip_iF(\bar\vphi^i_{AB},\phi^i_{AB})\ge 1-\eps/2}} \min_i {S_{\min}
(\rho^{\bar\vphi^i}_A)},
\end{align}
since $ \lambda_{\max}(\rho_B^{\bar\vphi^i})= \lambda_{\max}(\rho_A^{\bar\vphi^i})=S_{\min}(\rho_A^{\bar\vphi^i})$, with $\rho_B^{\bar\vphi^i}:= \tr_A({\bar\vphi^i})$ and $\rho_A^{\bar\vphi^i}:= \tr_B({\bar\vphi^i})$, because $\bar\vphi^i_{AB}$ is a pure state. To obtain the identity on the third line, we made use of 
the fact that $\Pi_{\bar{\sigma}_{ABZ}}=\sum_i \bar\vphi^i_{AB} \otimes \pi^i_Z$.
\end{proof}

The proof of Theorem \ref{thm_4} can be divided into the following two lemmas.
\begin{lemma}\label{asymp_direct}
For any bipartite state $\rho_{AB}$,
\be
D_A^\infty(\rho_{AB}) \ge \lim_{n \rightarrow \infty} \frac{1}{n} E_A(\rho_{AB}^{\otimes n}),
\ee
\end{lemma}
\begin{proof}
  Let $\eE=\{p_i,\phi^i_{AB}\}$ be an ensemble of pure states for
  $\rho_{AB}$ and $\eE_n=\{p_i^n, \phi^i_{A_nB_n}\}$ be an ensemble of
  pure states for $\rho_{AB}^{\otimes n}$. First of all, note that the
  pure states $\phi^i_{A_nB_n}$ need not be factorized. For this
  ensemble, define the tripartite state
\begin{equation}\label{sigma}
  \sigma_{ABZ}^{\eE_n} = \sum_i p_i^n \phi^i_{A_nB_n} \otimes
  \pi^{n,i}_Z \in {\cal{B}}\bigl(\sH_A^{\otimes n}\otimes
  \sH_B^{\otimes n}\otimes\sH_Z^{\otimes n}\bigr),
\end{equation}
where $\pi^{n,i}_Z = |i_n\rangle \langle i_n| \in
\states(\sH_Z^{\otimes n})$, with $\{ |i_n\rangle \}_i$ being an
orthonormal basis of $\sH_Z^{\otimes n}$.

From \reff{88} of Lemma \ref{thm_3} we have, for any given $\eps \ge 0$,
\begin{equation}
D_A(\rho_{AB}^{\otimes n}; \eps)\ge \max_{\eE_n}I^{A_n\leadsto B_nZ_n}_{0,\eps/2}
(\sigma_{ABZ}^{\eE_n}) - \Delta_n
\label{lbbd}
\end{equation} 
with $0\le \Delta_n \le 1$. We then have:



\begin{align}
D_A^\infty(\rho_{AB}):=& \lim_{\eps \rightarrow 0} \lim_{n \rightarrow \infty} \frac{1}{n} D_A( \rho_{AB}^{\otimes n} ; \eps),\nonumber\\
\ge &  \lim_{\eps \rightarrow 0} \lim_{n \rightarrow \infty} \frac{1}{n}
 \max_{\eE_n}I^{A_n\leadsto B_nZ_n}_{0,\eps/2}
(\sigma_{ABZ}^{\eE_n})\nonumber\\
\ge & \lim_{\eps \rightarrow 0} \lim_{n \rightarrow \infty} \frac{1}{n}
\max_{\eE}I^{A_n\leadsto B_nZ_n}_{0,\eps/2}\left((\sigma_{ABZ}^{\eE})^{\otimes n}\right)\nonumber\\
=&  \max_{\eE}\Bigl[I^{A\rightarrow BZ}(\sigma_{ABZ}^{\eE})\Bigr].
\label{long1}
\end{align}
The proof of~\reff{long1} can be found in Appendix~\ref{app:c}

From the definition of the state $\sigma_{ABZ}^{\eE}$ it follows that
for the ensemble $\eE=\{p_i, \phi^i_{AB}\}$,
\begin{equation}\label{idf}
  I^{A\rightarrow BZ}(\sigma_{ABZ}^{\eE}) = \sum_i p_i
  S(\rho^{\phi^i}_B),
\end{equation}
where $\rho^{\phi^i}_B = \tr_{AZ}\bigl(\sigma_{ABZ}^{\eE}\bigr)$. From
\reff{long1} and \reff{idf} we hence obtain
\begin{align}
  D_A^\infty(\rho_{AB}) &\ge  \max_{\eE} \sum_i p_i S(\rho^{\phi^i}_B)\nonumber\\
  &= E_A(\rho_{AB}).
\end{align}
The statement of the lemma can then be obtained by the usual blocking
argument.
\end{proof}

\begin{lemma}\label{asymp_converse}
For any bipartite state $\rho_{AB}$,
\begin{equation}
D_A^\infty(\rho_{AB}) \le \lim_{n \rightarrow \infty} \frac{1}{n} E_A(\rho_{AB}^{\otimes n}),
\end{equation}
\end{lemma}
\begin{proof}
From \reff{88} of Lemma \ref{thm_3} we have, for any given $\eps \ge 0$,
\begin{equation}
D_A(\rho_{AB}^{\otimes n};\eps)\le \max_{\eE_n}I^{A_n\to B_nZ_n}_{0,2\sqrt{\eps}}
(\sigma_{ABZ}^{\eE_n}),
\label{upbd}
\end{equation}
where the maximisation is over all possible pure state decompositions of the
satte $\rho_{AB}^{\otimes n}$. 

From Lemma 14 of \cite{qcap} we have the following inequality relating the smoothed zero-coherent information to the ordinary coherent information:
\begin{align}
I^{A_n\to B_nZ_n}_{0,2\sqrt{\eps}}(\sigma_{ABZ}^{\eE_n})
\le&  \frac{I^{A_n\to B_nZ_n}(\sigma_{ABZ}^{\eE_n})}{1 - \eps^{''}}\nonumber\\
& \quad + 
\frac{4\bigl( \eps^{''} \log\bigl(d_A^n d_{BZ}^n\bigr) 
+ 1\bigr)}{1-  \eps^{''}},
\end{align}
where $\eps' = 2\sqrt{\eps}$, $ \eps^{''}= 2 \sqrt{\eps'}$, $d_A^n= {\rm{dim }}
\sH_A^{\otimes n}$ and $d_{BZ}^n= {\rm{dim }} \bigl(\sH_B^{\otimes n}\otimes 
\sH_Z^{\otimes n}\bigr)$. Moreover, analogous to \reff{idf} we have
\be
I^{A_n\rightarrow B_nZ_n}(\sigma_{ABZ}^{\eE_n}) = \sum_i p_i^n S(\rho_{\phi^i}^{B_n}).
\ee
Hence,
\bea
D_A^\infty(\rho_{AB}) &\le & \lim_{n\rightarrow \infty} \frac{1}{n}\max_{\eE_n}
I^{A_n\rightarrow B_nZ_n}(\sigma_{ABZ}^{\eE_n})\nonumber\\
&=& \lim_{n\rightarrow \infty} \frac{1}{n}\max_{\eE_n}\sum_i p_i^n S(\rho_{\phi^i}^{B_n})\nonumber\\
&=& \lim_{n\rightarrow \infty} \frac{1}{n}E_A\bigl(\rho_{AB}^{\otimes n}\bigr)
\eea    
\end{proof}

\section{Discussion}

In this paper we evaluated the one-shot entanglement of assistance for
an arbitrary bipartite state $\rho_{AB}$. In doing this, we proved a
result, which is of interest on its own, namely a characterization of
the one-shot distillable entanglement of a bipartite pure state. This
result turned out to be stronger than what one obtains by simply
specializing the one-shot hashing bound, obtained in~\cite{distil}, to
pure states.

Further, we showed how our one-shot result yields the operational
interpretation of the asymptotic entanglement of assistance in the
asymptotic i.i.d. scenario. In this context, an interesting open
question is to find a one-shot analogue of the result
$E_A^\infty(\rho_{AB})=\min\{S(\rho_A),S(\rho_B)\}$ proved
in~\cite{assistance}.

\section*{Acknowledgments}

FB acknowledges support from the Program for Improvement of Research
Environment for Young Researchers from Special Coordination Funds for
Promoting Science and Technology (SCF) commissioned by the Ministry of
Education, Culture, Sports, Science and Technology (MEXT) of Japan.
ND acknowledges support from the European Community's Seventh
Framework Programme (FP7/2007-2013) under grant agreement number
213681. This work was done when FB was visiting the Statistical
Laboratory of the University of Cambridge.

\appendix

\section{Appendix A: optimality of rank-one measurements in~(\ref{eq:eoa-def})}\label{app:A}

Suppose in fact that the optimal assisting measurement at Charlie's is
given by the POVM $\{P^i_C\}_i$ (not necessarily rank-one). Then the
resulting shared state will be
$\sum_ip(i)\rho_{AB}^i\otimes\pi^i_X\otimes\pi^i_Y$, where
$p(i)\rho_{AB}^i= \Tr_C\left[(\openone_{AB}\otimes P^i_C)\
  \Psi_{ABC}\right]$, and $\pi^i$ is the shorthand notation for the
projector $|i\>\<i|$. In this form, the systems $X$ and $Y$, at
Alice's and Bob's side respectively, are classical registers carrying
the information about the outcome of Charlie's measurement.

Now, consider the situation where Charlie actually performs the
rank-one POVM $\{|\mu_i\>\<\mu_i|_C\}_{(i,\mu_i)}$, with
$\sum_{\mu_i}|\mu_i\>\<\mu_i|_C=P^i_C$, and communicates the double
index outcome $(i,\mu_i)$ to Alice and Bob. In this case, the shared
state between Alice and Bob can be written as
$\sum_{i,\mu_i}p(i,\mu_i)|\vphi^{(i,\mu_i)}\>\<\vphi^{(i,\mu_i)}|_{AB}\otimes\pi^i_X\otimes\pi^{\mu_i}_{X'}\otimes\pi^i_Y\otimes\pi^{\mu_i}_{Y'}$,
where $$p(i,\mu_i)
|\vphi^{(i,\mu_i)}\>\<\vphi^{(i,\mu_i)}|_{AB}=\Tr_C\left[\left(\openone_{AB}\otimes
    |\mu_i\>\<\mu_i|_C\right)\ \Psi_{ABC}\right].$$ It is easy to
verify that $\sum_{\mu_i}p(i,\mu_i)
|\vphi^{(i,\mu_i)}\>\<\vphi^{(i,\mu_i)}|_{AB}=p(i)\rho^i_{AB}$, so
that, in order to retrieve the optimal case, Alice and Bob simply have
to first perform a partial trace over the registers $X'$ and $Y'$,
respectively, and then proceed with the required LOCC
transformation. The partial trace can be effectively seen as a
coarse-graining of Charlie's measurement.





\section{Appendix B: proof of equation~\reff{long1}}\label{app:c}

Equation~\reff{long1} is proved by using Lemma~\ref{lemma:six}
and Lemma~\ref{lemma:seven}, given below. However, before stating and 
proving these lemmas,
we need to recall some definitions and notations extensively used in
the Quantum Information Spectrum
Approach~\cite{info-spect,hayashi-naga}. A fundamental quantity used in
this approach is the \emph{quantum spectral inf-divergence rate},
defined as follows~\cite{hayashi-naga}:
\begin{definition}[Spectral inf-divergence rate]
  Given a sequence of states $\hat\rho=\{\rho_n\}_{n=1}^\infty$, $\rho_n\in\states(\sH^{\otimes n})$, and a
  sequence of positive operators
  $\hat\sigma=\{\sigma_n\}_{n=1}^\infty$, with $\sigma_n\in {\cal B}(\sH^{\otimes n})$, the \emph{quantum spectral
  inf-divergence rate} is defined in terms of the difference
  operators $\Delta_n(\gamma) := \rho_n - 2^{n\gamma}\sigma_n$ as follows:
\begin{equation}
  \underline{D}(\hat\rho \| \hat\sigma) := \sup \left\{ \gamma :
    \liminf_{n\rightarrow \infty} \mathrm{Tr}\left[ \{ \Delta_n(\gamma)\ge 0\} \Delta_n(\gamma)\right] = 1 \right\}, \label{udiv}
\end{equation}
where the notation $\{X\ge 0\}$, for a self-adjoint operator $X$, is used to indicate the projector onto the non-negative eigenspace of $X$.
\end{definition}

\begin{lemma}\label{lemma:six}
For any given bipartite state $\rho_{AB}$, let $\eE$ denote a
pure-state ensemble decomposition, and let $\eE_n$ denote a pure-state ensemble
  decomposition of the state $\rho_{AB}^{\otimes n}$ . Then, using the notation of (\ref{sigma}), we have
\be
\lim_{\eps\to 0}\lim_{n\to\infty}\frac 1n\max_{\eE_n}I^{A_n\leadsto B_nZ_n}_{0,\eps}
(\sigma_{ABZ}^{\eE_n})
\ge \max_{\eE}\min_{\hat\nu_{BZ}}\underline{D}(\hat\sigma_{ABZ}^{\eE}\|
\hat\openone_A\otimes\hat\nu_{BZ}),\label{eq:rhs}
\ee
where $\hat\sigma_{ABZ}^{\eE}:=\left\{(\sigma_{ABZ}^{\eE})^{\otimes
    n}\right\}_{n\ge 1}$, $\hat\openone_A:=\{\openone_A^{\otimes
  n}\}_{n\ge 1}$, and
$\hat\nu_{BZ}:=\{\nu_{BZ}^n\in\states(\sH_B^{\otimes
  n}\otimes\sH_Z^{\otimes n})\}_{n\ge 1}$.
\end{lemma}

\begin{proof}
  Let $\bar\eE$ be the pure state ensemble decomposition of
  $\rho_{AB}$ for which the maximum on the r.h.s. of
  eq.~(\ref{eq:rhs}) is achieved. 
Since $\bar\eE$ is fixed, in the following, we drop
  the superscript $\bar\eE$ whenever no confusion arises, denoting
  $\sigma_{ABZ}^{\bar\eE}$ simply as $\sigma_{ABZ}$.

From the definition \reff{eq2} it follows that, for any fixed $\eps> 0$,
\begin{align}
  & \max_{\eE_n}I^{A_n\leadsto B_nZ_n}_{0,\eps}
  (\sigma_{ABZ}^{\eE_n})\nonumber\\
  =&\max_{\eE_n}\max_{\bar\sigma_{A_nB_nZ_n}^n\in\B_{\textrm{qc}} (\sigma_{ABZ}^{\eE_n};\eps)}\min_{\nu_{B_nZ_n}^n}S_0(\bar\sigma_{A_nB_nZ_n}^{\eE_n}\|\openone_A^{\otimes n}\otimes\nu_{B_nZ_n}^n)\nonumber\\
  \ge&\max_{\eE}\max_{\bar\sigma_{A_nB_nZ_n}^n\in\B_{\textrm{qc}} ((\sigma_{ABZ}^{\eE})^{\otimes n};\eps)}\min_{\nu_{B_nZ_n}^n}S_0(\bar\sigma_{A_nB_nZ_n}^n\|\openone_A^{\otimes n}\otimes\nu_{B_nZ_n}^n)\nonumber\\
  \ge&\max_{\bar\sigma_{A_nB_nZ_n}^n\in\B_{\textrm{qc}}
    (\sigma_{ABZ}^{\otimes
      n};\eps)}\min_{\nu_{B_nZ_n}^n}S_0(\bar\sigma_{A_nB_nZ_n}^n\|\openone_A^{\otimes
    n}\otimes\nu_{B_nZ_n}^n).\label{eq:here2}
\end{align}

For each $\nu_{B_nZ_n}^n$ and any $\gamma\in\mathbb{R}$, define the projector \begin{equation}
  P_n^\gamma\equiv P_n^\gamma(\nu_{B_nZ_n}^n):=\{\sigma_{ABZ}^{\otimes n}- 2^{n\gamma}(\openone_A^{\otimes n}\otimes\nu_{B_nZ_n}^n) \ge 0\}.
\end{equation}
Since the operator $\bar\sigma_{A_nB_nZ_n}^n$ in~(\ref{eq:here2}) is a
qc operator, it is clear that the minimization over $\nu_{B_nZ_n}^n$
in~(\ref{eq:here2}) can be restricted to states diagonal in the basis
chosen in representing qc operators. Consequently, also $P_n^\gamma$
has the same qc structure.

Next, let us denote by $\hat\sigma_{ABZ}$ the i.i.d. sequence of
states $\{\sigma_{ABZ}^{\otimes n}\}_{n\ge 1}$. For any sequence
$\hat\nu_{BZ}:=\{\nu_{B_nZ_n}^n\}_{n\ge 1}$, fix $\delta>0$ and choose
$\gamma\equiv\gamma(\hat\nu_{BZ}):=
\underline{D}(\hat\sigma_{ABZ}\|\hat\openone_A\otimes\hat\nu_{BZ})
-\delta$. Then it follows from the definition~(\ref{udiv}) that, for
$n$ large enough,
\begin{equation}
  \Tr\left[P_n^\gamma\
    \sigma_{ABZ}^{\otimes n}\right]\ge 1-\frac{\eps^2}4,
\end{equation}
for any $\eps>0$. Further, define
\begin{equation}
\omega_{A_nB_nZ_n}^{n,\gamma}\equiv
\omega_{A_nB_nZ_n}^{n,\gamma}(\nu_{B_nZ_n}^n):=\frac{{\sqrt{P_n^\gamma}}
\sigma_{ABZ}^{\otimes n} {\sqrt{P_n^\gamma}}}{\Tr\left[P_n^\gamma
\sigma_{ABZ}^{\otimes n}\right]},
\end{equation}
which, by Lemma~\ref{lemma:accessory}, is clearly in $\B_{\textrm{qc}}(\sigma_{ABZ}^{\otimes n};\eps)$, the qc-ball around the state $\sigma_{ABZ}^{\otimes n}$,defined by \reff{bqc2}.

Then, using the fact that $\Pi_{\omega_{A_nB_nZ_n}^{n,\gamma}} \le P_n^\gamma$, and Lemma~2 of~\cite{nila}, we have, for any fixed $\eps>0$,
\begin{eqnarray}
&& \lim_{n\to\infty}\frac 1n\,\{\textrm{r.h.s. of (\ref{eq:here2})}\}\nonumber\\
&\ge & \lim_{n\to\infty}\frac 1n\min_{\nu_{B_nZ_n}^n} 
S_0(\omega^{n,\gamma}_{R_nA_n}\|\openone_A^{\otimes n}\otimes \nu^n_{B_nZ_n})\nonumber\\
&= &\lim_{n\to\infty}\frac 1n\min_{ \nu^n_{B_nZ_n}}\left\{-\log\Tr\left[\Pi_{\omega^{n,\gamma}_{A_nB_nZ_n}}(\openone_A^{\otimes n}\otimes \nu_{B_nZ_n}^n)\right]\right\}\nonumber\\
 &\ge & \lim_{n\to\infty}\frac 1n\min_{ \nu^n_{B_nZ_n}}\left\{ -\log\Tr\left[P_n^\gamma(\openone_A^{\otimes n}\otimes  \nu^n_{B_nZ_n})\right]\right\}\nonumber\\
 &\ge &\min_{\hat\nu_{BZ}} \gamma(\hat\nu_{BZ})\nonumber\\
&=& \min_{\hat\nu_{BZ}}\underline{D}(\hat\sigma_{ABZ}\|\hat\openone_A\otimes
\hat\nu_{BZ})-\delta\nonumber\\
&=& \max_{\eE}\min_{\hat\nu_{BZ}}\underline{D}(\hat\sigma_{ABZ}^\eE\|\hat\openone_A\otimes \hat\nu_{BZ})-\delta
\end{eqnarray}
Since this holds for any arbitrary $\delta>0$, it yields the required
inequality~(\ref{eq:rhs}) in the limit $\eps\to 0$.
\end{proof}

We also use the following lemma from \cite{cost}, which employs the
Generalized Stein's Lemma~\cite{brandao-plenio} and Lemma~4
of~\cite{qcap}. We include its proof for the sake of completeness.
\begin{lemma}\label{lemma:seven}
For any given bipartite state $\rho_{AR}$,
\begin{equation}
\min_{\hat\sigma_R}\underline{D}(\hat\rho_{AR}\|\hat\openone_A\otimes \hat\sigma_R)=S(\rho_{AR}\|\openone_A\otimes \rho_R),
\end{equation}
where $\hat\rho_{AR}=\{\rho_{AR}^{\otimes n}\}_{n\ge 1}$, $\rho_R = \Tr_A \rho_{AR}$, $\hat\sigma_R:=\{\sigma_{R_n}^n\in\states(\sH_R^{\otimes n})\}_{n\ge 1}$, and $\hat\openone_A:=\{\openone_A^{\otimes n}\}_{n\ge 1}$.
\end{lemma}

\begin{proof}Consider the family of sets $\mM:=\{\mM_n\}_{n\ge 1}$
\begin{equation}\label{eq:defsets}
  \mM_n:=\left\{\tau_{A_n}^n\otimes\sigma_{R_n}^n\in\states(\sH_A^{\otimes n} \otimes \sH_R^{\otimes n})\right\},
\end{equation}
such that $\tau_{A_n}^n:=(\openone_A/d_A)^{\otimes n}$. For this family, the Generalized Stein's Lemma~(Proposition III.1 of~\cite{brandao-plenio}) holds.

More precisely, for a given bipartite state $\rho_{AR}$, let us define
\begin{equation}
  S_\mM^\infty(\rho_{AR}):=\lim_{n\to\infty}\frac 1nS_{\mM_n}(\rho_{AR}^{\otimes n}),
\end{equation}
with $S_{\mM_n}(\rho_{AR}^{\otimes n}):=\min_{\omega^n_{A_nR_n}\in\mM_n}S(\rho^{\otimes n}_{AR}\|\omega^n_{A_nR_n})$, and
$\Delta_n(\gamma) = \rho_{AR}^{\otimes n} - 2^{n\gamma}\omega^n_{A_nR_n}$. From the Generalized Stein's Lemma~\cite{brandao-plenio} it follows that, for $\gamma>S^\infty_\mM(\rho_{AR})$, 
\begin{equation}
  \lim_{n\to\infty}\min_{\omega^n_{A_nR_n}\in\mM_n}\Tr\left[\{\Delta_n(\gamma)\ge 0\}\Delta_n(\gamma)\right]=0,
\end{equation}
implying that
$\min_{\hat\omega_{AR}\in\mM}\underline{D}(\hat\rho_{AR}\|\hat\omega_{AR})\le S^\infty_\mM(\rho_{AR})$.
On the other hand, for $\gamma<S^\infty_\mM(\rho_{AR})$,
\begin{equation}
  \label{eq:36}
  \lim_{n\to\infty}\min_{\omega^n_{A_nR_n}\in\mM_n}\Tr\left[\{\Delta_n(\gamma)\ge 0\}\Delta_n(\gamma)\right]=1,
\end{equation}
implying that $\min_{\hat\omega_{AR}\in\mM}\underline{D}(\hat\rho_{AR}\|\hat\omega_{AR})\ge S^\infty_\mM(\rho_{AR})$.
Hence $$\min_{\hat\omega_{AR}\in\mM}\underline{D}(\hat\rho_{AR}\|\hat\omega_{AR})= S^\infty_\mM(\rho_{AR}).$$

Finally, by noticing that, due to the definition~(\ref{eq:defsets}) of $\mM$,
\begin{equation}
\begin{split} \min_{\hat\omega_{AR}\in\mM}&\underline{D}(\hat\rho_{AR}\|\hat\omega_{AR})\\
=\min_{\hat\sigma_R}&\underline{D}(\hat\rho_{AR}\|\hat\openone_A\otimes 
\hat\sigma_R)+\log d_A,
\end{split}
\end{equation}
and that, due to Lemma~4 in~\cite{qcap},
\begin{equation}
  S^\infty_\mM(\rho_{AR})=S(\rho_{AR}\|\openone_A\otimes \rho_R)+\log d_A,
\end{equation}
we obtain the statement of the lemma.
\end{proof}
From Lemma~\ref{lemma:six} and Lemma~\ref{lemma:seven} we conclude that 
\bea
\lim_{\eps\to 0}\lim_{n\to\infty}\frac 1n\max_{\eE_n}I^{A_n\leadsto B_nZ_n}_{0,\eps}
(\sigma_{ABZ}^{\eE_n})
&\ge & \max_{\eE}\min_{\hat\nu_{BZ}}\underline{D}(\hat\sigma_{ABZ}^{\eE}\|
\hat\openone_A\otimes\hat\nu_{BZ})\nonumber\\
&= & \max_{\eE}S(\sigma_{ABZ}^{\eE}\|
\openone_A\otimes\sigma_{BZ}^{\eE})\nonumber\\
&=&  \max_{\eE}\Bigl[I^{A\rightarrow BZ}(\sigma_{ABZ}^{\eE})\Bigr],
\eea
where $\sigma_{BZ}^{\eE}=\Tr_A\sigma_{ABZ}^{\eE}$. Thus \reff{long1} is proved.

\section*{Biographies}

Francesco Buscemi received the Ph.D. in Physics from the University of
Pavia, Italy, in 2006. From 2008 to 2009, he was Research Associate
at the Statistical Laboratory of the University of Cambridge. From
2009 he joined Nagoya University, Japan, as Designated Associate
Professor at the Institute of Advanced Research and joint member of
the Graduate School of Information Science.\bigskip

\noindent Nilanjana Datta received a Ph.D. degree from ETH Zurich, Switzerland,
in 1996.  From 1997 to 2000, she was a postdoctoral researcher at the
Dublin Institute of Advanced Studies, C.N.R.S. Marseille, and EPFL in
Lausanne. In 2001 she joined the University of Cambridge, as a
Lecturer in Mathematics of Pembroke College, and a member of the
Statistical Laboratory, in the Centre for Mathematical Sciences. She
is currently an Affiliated Lecturer of the Faculty of Mathematics,
University of Cambridge, and a Fellow of Pembroke College.

\end{document}